\documentclass[10pt]{article}
\usepackage{amsmath}
\usepackage{mathabx}
\usepackage{graphicx}
\usepackage{multirow}
\usepackage{color}
\usepackage{orcidlink}
\usepackage{hyperref}
\hypersetup{colorlinks=true, linkcolor=blue, citecolor=blue, urlcolor=blue}
\usepackage{subcaption}
\begin{document}
\baselineskip=16pt

\begin{center}
\Large{Shadow properties and orbital dynamics around an effective quantum-modified black hole surrounded by quintessential dark energy}
\end{center}

\vspace{0.3cm}

\begin{center}

Ahmad Al-Badawi \orcidlink{0000-0002-3127-3453}\\
Department of Physics, Al-Hussein Bin Talal University, P. O. Box: 20, 71111,
Ma'an, Jordan. \\
e-mail: ahmadbadawi@ahu.edu.jo (Corresp. author)\\

Faizuddin Ahmed\orcidlink{0000-0003-2196-9622}\\Department of Physics, University of Science \& Technology Meghalaya, Ri-Bhoi, Meghalaya, 793101, India\\
e-mail: faizuddinahmed15@gmail.com\\

Tursunali Xamidov\orcidlink{0009-0008-3081-9646}\,\,\,,\,\,\,Sanjar Shaymatov\orcidlink{0000-0002-5229-7657}\footnote[1]{Other affiliations: Institute for Theoretical Physics and Cosmology, Zhejiang University of Technology, Hangzhou 310023, China\\

University of Tashkent for Applied Sciences, Str. Gavhar 1, Tashkent 100149, Uzbekistan\\

Western Caspian University, Baku AZ1001, Azerbaijan} \\ Institute of Fundamental and Applied Research, National Research University TIIAME, Kori Niyoziy 39, Tashkent 100000, Uzbekistan\\ e-mail: xamidovtursunali@gmail.com\,\,;\,\,\,e-mail: sanjar@astrin.uz\\

\.{I}zzet Sakall{\i}\orcidlink{0000-0001-7827-9476}\\
Physics Department, Eastern Mediterranean University, Famagusta 99628, North Cyprus via Mersin 10, Turkey\\
e-mail: izzet.sakalli@emu.edu.tr

\end{center}

\vspace{0.3cm}

\begin{abstract}

In this study, we investigate black holes (BHs) surrounded by a quintessence field (QF) within the framework of effective quantum gravity (EQG). We analyze the spacetime metric characterized by quantum correction parameter $\xi$ and quintessence parameters $(c,w)$, revealing a rich three-horizon structure whose properties depend on both quantum effects and dark energy. Our main focus is on understanding the observational signatures and dynamical behavior of these modified BHs. We derive analytical expressions for the photon sphere and shadow radius for various values of the state parameter, finding that quintessence fields increase the shadow radius while quantum corrections decrease it-a distinctive interplay that creates potentially observable effects. Using Event Horizon Telescope data from M87*, we establish constraints on the quantum correction parameter, showing that the allowable range for $\xi$ increases with the quintessence parameter $c$. We conduct a comprehensive analysis of null and timelike geodesics, demonstrating how quantum corrections enhance interactions between photons and the gravitational field while modifying the energy, angular momentum, and stability properties of massive particle orbits. Our investigation extends to periodic orbits characterized by zoom-whirl behavior, finding that quintessence allows such orbits to occur at lower energies compared to purely quantum-corrected BHs. We further analyze scalar perturbations and their effective potentials, which reveal how both quantum and quintessence effects shape the response characteristics of these BHs. Throughout our study, we find that quantum corrections and quintessence frequently produce counteracting effects on observables that could provide simultaneous tests of quantum gravity theories and dark energy models through future high-precision astronomical observations.
\end{abstract}


\newpage

\section{Introduction}

BHs stand as extraordinary predictions of Einstein's general relativity, representing regions where gravity becomes so extreme that nothing can escape. These objects have transitioned from theoretical constructs to observational realities through gravitational wave detections by LIGO-Virgo \cite{isz01,isz02} and direct imaging by the Event Horizon Telescope \cite{isz03,isz04}. Despite these breakthroughs, classical general relativity remains inadequate at the quantum level, where singularities indicate a breakdown of the theory.

EQG has emerged as a pragmatic approach to this problem, treating quantum gravity effects as corrections to classical spacetime metrics without requiring a complete quantum theory of gravity \cite{isz05,isz06}. This framework preserves mathematical tractability while incorporating quantum effects through parameters that can potentially be constrained by observations \cite{isz07,isz08}. Recent work has yielded several quantum-corrected BH solutions that exhibit distinct deviations from their classical counterparts in both geometric and thermodynamic properties \cite{isz09}. Simultaneously, cosmological observations have established the accelerated expansion of our universe, attributed to mysterious dark energy \cite{isz10,isz11}. Among various theoretical models explaining this phenomenon, QFs have received significant attention. Quintessence is a dynamical scalar field with negative pressure characterized by a state parameter $w$ (typically within $-1 < w < -1/3$) that can account for cosmic acceleration \cite{isz12,isz13,isz14}. Unlike the cosmological constant ($w = -1$), quintessence offers a time-dependent equation of state, providing greater flexibility in explaining the evolution of dark energy \cite{isz15}. It is also worth noting that the incorporation of quintessence into BH physics, pioneered by Kiselev \cite{isz16}, has generated numerous studies examining how cosmic dark energy influences local spacetime geometry around BHs \cite{isz17,isz18,isz18a,isz18b}. 

The shadow of a BH-its apparent outline as seen by distant observers-provides a crucial observational signature for testing modified gravity theories \cite{isz19,isz20}. Recent EHT observations have begun placing constraints on alternative theories, with future advancements promising even more stringent tests \cite{isz03}. In quantum-corrected space-times surrounded by quintessence, shadow properties acquire distinctive modifications that potentially allow for observational discrimination between competing theories \cite{isz22}. Geodesic motion around BHs offers another powerful probe of spacetime geometry. Null geodesics (followed by photons) determine not only shadow characteristics but also lensing effects, while timelike geodesics (followed by massive particles) govern orbital dynamics of matter around BHs \cite{isz23,isz24}. In EQG with quintessence, these geodesic equations contain additional terms leading to potentially observable deviations in particle trajectories \cite{isz25}. The innermost stable circular orbit (ISCO)-the smallest radius permitting stable circular motion-marks the inner edge of thin accretion disks and significantly influences BH the electromagnetic signatures \cite{isz26,isz26a,isz26b}. Both quantum corrections and QFs modify the ISCO radius, affecting the observational features of accreting systems \cite{isz27}. Beyond ISCO, bound orbits around modified BHs exhibit complex patterns, including periodic orbits characterized by zoom-whirl behavior, where particles alternate between quasi-circular and highly eccentric phases \cite{isz28,isz29}. Periodic orbits have gained significant attention since the detection of GWs \cite{isz01,isz02} due to their importance in understanding the sources of these waves, especially compact binary objects and extreme-mass-ratio inspiral (EMRI) systems. The tightly bound orbits in the inspiral phase of these systems can be understood as transitions between successive periodic orbits, which makes them crucial for gravitational wave emission. In \cite{isz28} a classification system using three integers ($z$, $\omega$, $v$) was introduced to categorize periodic orbits of massive particles around black holes. For fixed angular momentum and energy, the triplet ($z$, $\omega$, $v$) defines a rational number $q$, which characterizes a closed periodic orbit. In recent years, there has since been increased research activity addressing periodic orbits within various gravity scenarios \cite{Levin09,Levin10,Tu23,Mustapha20,Jiang24}. These dynamics have implications for gravitational wave signals from extreme mass-ratio inspirals and stellar distributions near supermassive BHs \cite{isz30}.

The stability of modified BH solutions against perturbations constitutes another crucial aspect of their viability. Scalar perturbations serve as valuable test cases for understanding how disturbances propagate in quantum-corrected space-times with quintessence \cite{isz31,isz32}. The effective potential governing scalar field propagation encodes information about the background geometry, determining quasinormal modes and greybody factors that characterize the BH's response to external perturbations and its Hawking radiation spectrum \cite{isz33,isz34,isz34x3}. The Lyapunov exponent, quantifying the rate at which nearby geodesics diverge, measures the chaotic properties of orbital motion and connects to the quasinormal mode spectrum through the correspondence between high-frequency modes and unstable circular null geodesics \cite{isz35}. In quantum-modified spac-etimes with quintessence, modified Lyapunov exponents potentially offer another observational channel for testing these theories \cite{isz36}.


Despite extensive research on quantum-modified black holes (BHs) and quintessence fields (QFs) surrounding BHs separately, their combined effects have received comparatively less attention. This gap motivates our investigation into effective quantum-corrected BHs immersed in QFs, recognizing that both quantum effects near the horizon and cosmological dark energy fields are likely to coexist in realistic scenarios, making their joint consideration essential for accurate modeling of observable phenomena. In this study, we analyze EQG BH solutions characterized by the quantum-correction parameter $\xi$ and quintessence parameters $(c, w)$, where $c$ is the normalization factor and $w$ is the equation of state parameter. Our analysis includes investigating the horizon structure, deriving analytical expressions for shadow radii, examining null and time-like geodesics, exploring periodic orbits, and studying the dynamics of spin-0 massless scalar field perturbations along with their effective potentials. Our results show that both quantum corrections and QFs significantly influence BH shadows, geodesic structure, and perturbative behavior, often in opposing ways-while quintessence tends to enlarge the shadow and orbital radii, quantum corrections generally reduce these quantities, leading to a rich phenomenology that deviates significantly from both classical predictions and singly-modified theories.

The paper is organized as follows. In Section \ref{sec2}, we present the metric for effective quantum-corrected BHs surrounded by QFs and analyze its horizon structure. Section \ref{sec3} explores shadow properties and establishes observational constraints based on EHT data. Section \ref{sec4} investigates geodesic motions of test particles, including associated forces, effective potentials, and Lyapunov exponents. Section \ref{sec5} examines periodic orbits and their classification. Section \ref{sec6} analyzes scalar perturbations and their effective potentials. Finally, in Sec. \ref{sec7}, we summarize our findings and discuss their implications for future theoretical and observational studies.

\section{Effective quantum corrected BH surrounded by QF} \label{sec2}

Ref. \cite{CZ} tackles the issue of maintaining covariance in the setting of the spherically symmetric sector of vacuum gravity. By keeping the theory's kinematical variables and the classical form of the vector constraints, the study introduces an arbitrary effective Hamiltonian constraint as well as a freely selected function for building the effective metric. As with classical theory, it is assumed that a Dirac observable expressing BH mass exists. In light of these assumptions, the authors of \cite{CZ} derive equations that ensure spacetime covariance based on these assumptions. These requirements establish links between the effective Hamiltonian, the Dirac observable for BH mass, and the free function. As a result, two distinct effective quantum-corrected metrics are obtained, with classical constraints recovering when quantum parameters are set to zero. We will focus on the first type BH model  surrounded by QF.  The metric of the  space-time of a static, spherically symmetric EQG BH with QF  is described by the following line-element \cite{CZ, VVK}
\begin{eqnarray}
&&ds^2=-\mathcal{F}(r)\,dt^2+\frac{dr^2}{\mathcal{F}(r)}+r^2\,\left(d\theta^2+\sin ^2 \theta \,d \phi^2\right),\nonumber\\
&&\mathcal{F}(r)=f(r)+\frac{\xi^2}{r^2}\,f^2(r),\nonumber\\
&&f(r)=1-\frac{2\,M}{r}-\frac{c}{r^{3\,w+1}},\quad -1 < w < -1/3,
\label{aa1}
\end{eqnarray}
with $\xi$ is the quantum correction parameter, $M$ denotes the ADM mass, $w$ represents the state parameter of the quintessence matter with $c$ serves as a positive normalization factor. In the limit where $c = 0$, the metric (\ref{aa1}) reduces to the BH metric (Model-I) discussed in Refs. \cite{CZ}. Additionally, in the limit where $\xi = 0$, corresponding to the absence of quantum corrections, the metric (\ref{aa1}) simplifies to the Kiselev BH metric, as presented in Ref. \cite{VVK}, which has been extensively studied in the literature. In this work, we aim to investigate the four-dimensional BH solution described in (\ref{aa1}) that incorporates the combined effects of quantum corrections and the QF. 

\begin{figure}[ht!]
    \centering
    \includegraphics[width=0.45\linewidth]{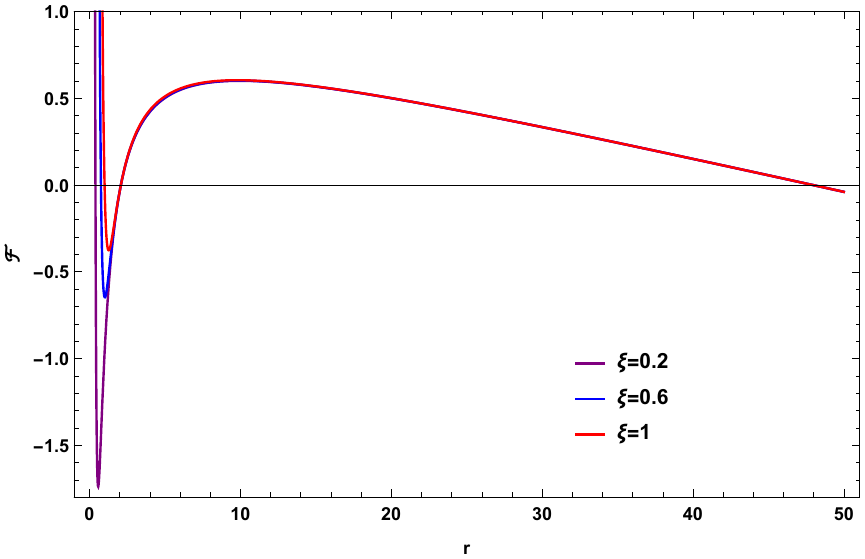}\quad\quad
    \includegraphics[width=0.45\linewidth]{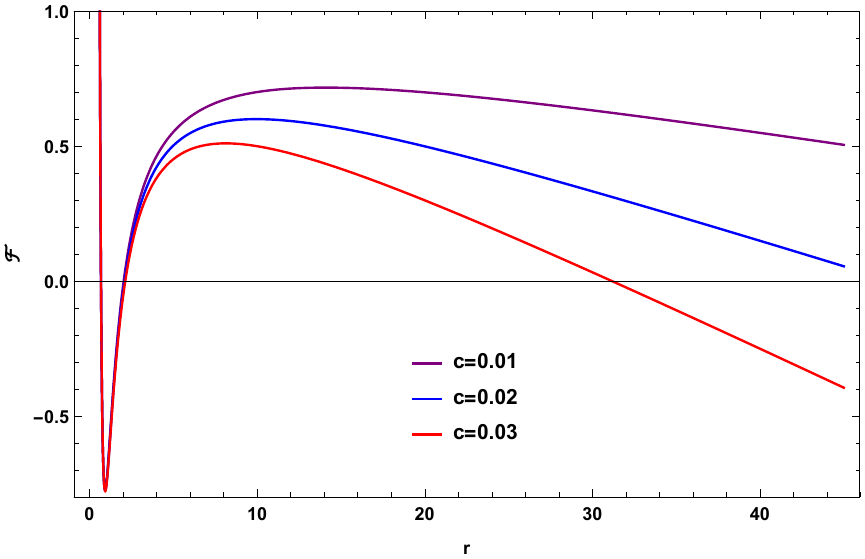}\\
    \includegraphics[width=0.45\linewidth]{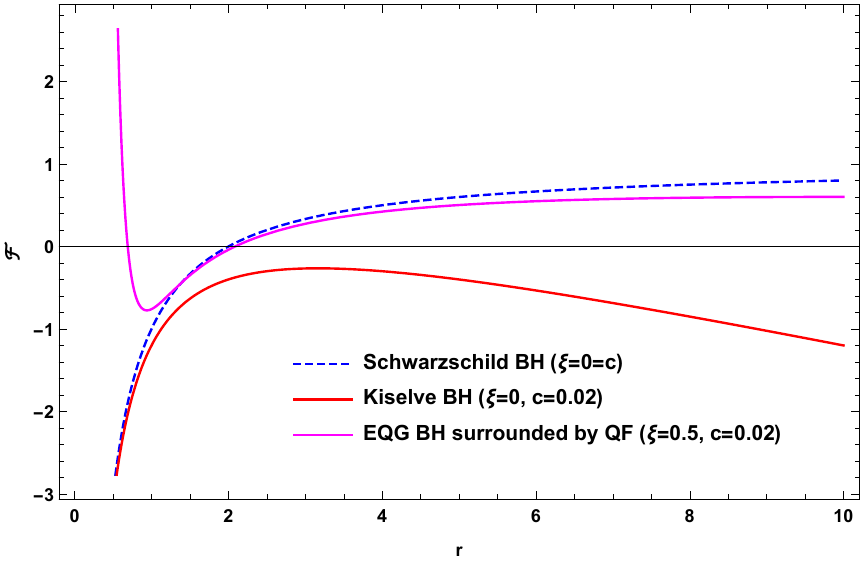}
    \caption{Plot of the metric function $f(r)$ versus $r$ for different values of quantum correction parameter $\xi$ and the arbitrary constant $c$. Here, $M=1$, $w=-2/3$. Top left panel: $c=0.02$; Top right panel: $\xi=0.5$.}
    \label{fig:metric-function-1}
\end{figure}

To visualize the metric functions  (\ref{aa1}), we display it as a function of $r$, as illustrated in Fig. \ref{fig:metric-function-1}. The figure shows that  EQG BH with
QF can have three horizons, namely the Cauchy radius $r_-$, event horizon $r_+$ and cosmological horizon $r_c$. We will now explore the horizon structure of the EQG BH with QF, finding the zeros of $\mathcal{F}(r_\text{h}) = 0$, which implies that
\begin{equation}
  \left(1-\frac{2\,M}{r_\text{h}}-\frac{c}{r^{3\,w+1}_\text{h}}\right)+\frac{\xi^2}{r^2_\text{h}}\left( 1-\frac{2\,M}{r_\text{h}}-\frac{c}{r^{3\,w+1}_\text{h}} \right)^2=0. \label{cc9} 
\end{equation}
The solution of Eq. (\ref{cc9}) is dependent on the choice of $w$. To solve Eq. (\ref{cc9}), we will focus on the choice $w=-2/3$. Thus, the analytical solutions for the horizons are
\begin{equation*}
    r_{+}=\frac{1-\sqrt{1-8\,c\,M}}{2\,c},\hspace{1cm}r_{c}=\frac{1+\sqrt{1-8\,c\,M}}{2\,c}\end{equation*}
\begin{equation}
r_{-}=\frac{c\,\xi^{2}}{3} +\frac{2^{1/3}\,\xi ^{2}\left( 3-c^{2}\,\xi^{2}\right) }{3\left( Y+3\,\sqrt{3\,X}\right) ^{1/3}}-\frac{1}{2^{1/3}\,3}\left( Y+3\,\sqrt{3\,X}\right) ^{1/3}, \label{horz1}
\end{equation}
where we set the following:
\begin{eqnarray}
Y=-54\,M\,\xi ^{2}+9\,c\,\xi ^{4}-2\,c^{3}\,\xi ^{6},\quad\quad 
X=108\,M\,\xi ^{4}+4\,\xi ^{6}-36\,c\,M\,\xi ^{6}-c^{2}\,\xi ^{8}+8\,c^{3}\,M\,\xi ^{8}.\label{variable}
\end{eqnarray}

\begin{figure}
\begin{center}
\includegraphics[scale=0.47]{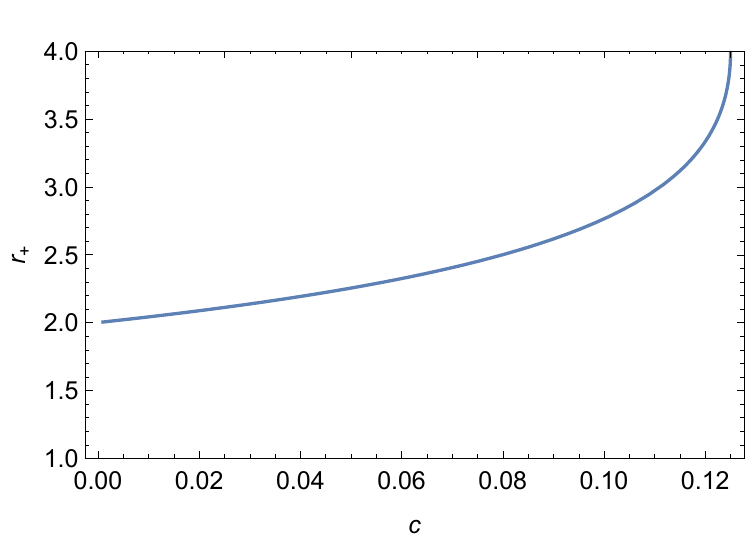}
\includegraphics[scale=0.47]{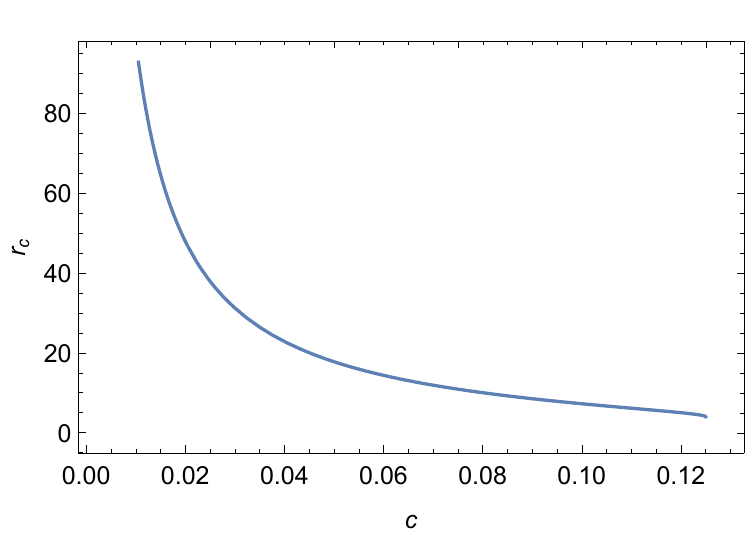}
\includegraphics[scale=0.47]{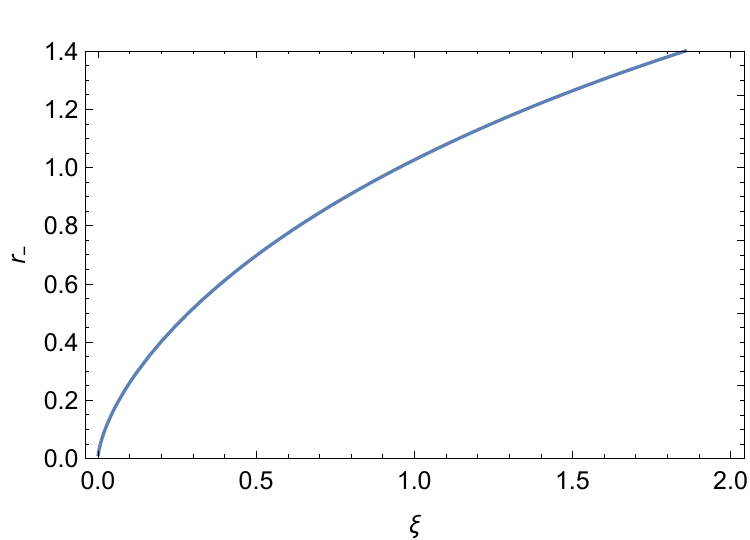}
\end{center}
\caption{Plot of the three horizons of the EQG BH with
QF with $w=-2/3$ and $M=1$. }\label{figa9}
\end{figure}

We note that, event horizon $r_+$ and cosmological horizon $r_c$ do not depend on quantum correction parameter $\xi$. Figure \ref{figa9} provides a detailed analysis of the three horizons. The graph shows that increasing the parameter $c$ leads to an increase in the event horizon (left panel). However, when the value of the  parameter $c$ increases, the cosmological horizon decreases (see middle panel). Finally, when the quantum parameter increases, so does the Cauchy horizon (right panel). It should be noted that the parameter $c$ has no significant effect on the Cauchy horizon.

\section{BH shadow and observational constraints} \label{sec3}

Many methods have been proposed to calculate the shadow radius of a spherical BH; for a recent overview, see \cite{shad99}. For a static spherically symmetric metric, the radius $r_{ph}$ of the photon ring around the BH is given by the equation \cite{shad99} \begin{equation}
    r\,D'(r)=D(r),
\end{equation}
where $D=\sqrt{\mathcal{F}}$. 

Considering  the metric function (\ref{aa1}), then the equation for $r_{ph}$ is 
\begin{equation}
\left\{2\,(3\,M-r)\,r^{3w}+3\,c\,(1+w)\right\}\left\{2\,c\,\xi^2-r^{3w}\,\left(r^3+2\,(r-2\,M)\,\xi^2\right)\right\}=0. \label{eps1}
\end{equation} 
When $c=0=w$ or $\xi=0$, then the equation (\ref{eps1}) reduces to $3\,M$. Again, the solution of Eq. (\ref{eps1}) depends on the choice of $w$. Analytically, we obtain the photon sphere for different selections of the state parameter $w$ as follows:
 
\begin{eqnarray}
&&(i)\,\,\mbox{for}\,\, w=-1/3, \,\, Eq. (\ref{eps1}) \Rightarrow \left(3M+(c-1)r\right)\left(r^3+2(r(1-c)-2M)\xi^2\right)=0, \,\,r_{ph}=\frac{3M}{1-c}.  
 \\
&&(ii)\,\,\mbox{for}\, \, w=-2/3,\,\, Eq. (\ref{eps1}) \Rightarrow
(6M+(cr-2)r)(-r^3+2(2M+r(cr-1))\xi^2)=0 ,\,\,r_{ph}=\frac{1-\sqrt{1-6cM}}{c}.\\
&&(iii)\,\,\mbox{for}\,\,  w=-1,\,\,  Eq. (\ref{eps1}) \Rightarrow ((3M-r)r^{-3})(2c\xi^2-r^{-3}(r^3+2(r-2M)\xi^2))=0, \,\,r_{ph}=3M . \label{bb16c}
\end{eqnarray}
It is important to highlight that the above solutions of the $r_{ph}$ constitute  the only real solutions. As a result, the photon sphere is determined solely by the quintessence parameters $w$ and $c$. Using photon radius, we can now compute the shadow radius $R_s$ observed by an observer at infinity as \begin{equation}
    R_s=\frac{r_{ph}}{\sqrt{f(r_{ph})}} .\label{shadeq1}
\end{equation}
Based on the metric (\ref{aa1}), we provide analytical solutions for the shadow radius of the EQG BH with QF as \begin{eqnarray}
  &&(i)\,\,\mbox{for}\,\, w=-1/3, \,\,R_{s}= \frac{27M}{(1-c)\sqrt{27(1-c)+\frac{(c-1)^4\xi^2}{M^2}}}     
 .\\
&&(ii)\,\,\mbox{for}\, \, w=-2/3, \,\,R_{s}= \frac{1-\sqrt{1-6cM}}{c\sqrt{\sqrt{1-6cM}+\frac{2cM}{\sqrt{1-6cM}-1}+\frac{c^2(4cM+\sqrt{1-6cM}-1)^2\xi^2}{(\sqrt{1-6cM}-1)^2}}}.\\
&&(iii)\,\,\mbox{for}\,\, w=-1,\,\,R_{s}=   \frac{27M}{\sqrt{(9cM-1)[(9cM-1)\frac{\xi^2}{M^2}-27]}}.     \label{bb16cc}
\end{eqnarray}
If $c=0$, that is, without QF surrounding the BH spacetime, we recover the EQG BH shadow radius $R_s=\frac{27M}{\sqrt{27+\xi^2/M^2}}$ \cite{Yu-Heng}.\\ 
In \cite{isz08}, the constraint on the quantum correction parameter $\xi$ in Eq. (\ref{aa1}) was explored using Sgr A* data which
is \begin{equation}
    0\leq \xi \lesssim 2.866 \, (1 \sigma),
\end{equation}
whereas the constraint on quantum correction parameter $\xi$  from data of M87* was considered in \cite{Yu-Heng}, which
is  \begin{equation}
    0\leq \xi \lesssim 2.304 \, (1 \sigma).
\end{equation}
In this section, we consider the constraint on $\xi$ based on the data of  M87*. The diameter of the shadow normalized by the gravitational radius of  M87* is \begin{equation}
    d_{M87}=\frac{\theta_{sh}D_{M87}}{M_{M87}}=11.0\pm 1.5.
\end{equation}
Thus, the shadow radius $R_s$ should satisfy the following relation within $(1 \sigma)$ \begin{equation}
    4.75\lesssim R_s \lesssim 6.25,
\end{equation}
this constraint leads to the following quantum correction parameter,
\begin{eqnarray}
  &&(i)\,\,\mbox{for}\,\, w=-1/3, \,\,
 \,\,0\leq \xi \leq \sqrt{\frac{(27/4.75)^2-27(1-c)^3}{(1-c)^6}},    
 \\
&&(ii)\,\,\mbox{for}\, \, w=-2/3,\,\,
  \,\,0\leq \xi \leq \sqrt{\frac{(A-1)^4\left((1/4.75)^2(A-1)^2 -c^2 A-(2c^3/(A-1))\right)}{c^4(A-1+4c)^2}}, A=\sqrt{1-6c} \\
&&(iii)\,\,\mbox{for}\,\, w=-1,\,\,
 \,\,0\leq \xi \leq \sqrt{\frac{(27/4.75)^2-27(1-9c)^2}{(1-9c)^3}}, \label{bb15}
\end{eqnarray}
 
\begin{figure}
    \centering
    \includegraphics[width=0.45\linewidth]{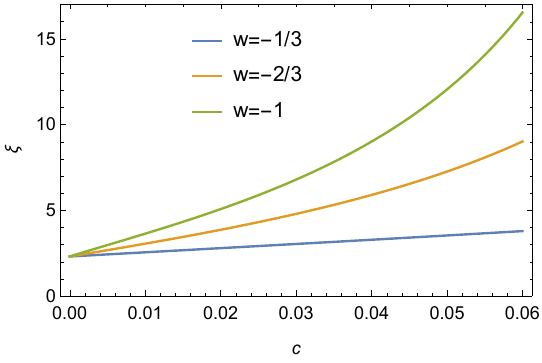}
    \caption{Variation of the permissible quantum corrected parameter range with the $c$ parameter for various state parameter $w$.}
    \label{figa8}
\end{figure}

\begin{figure}[ht!]
    \centering
    \includegraphics[width=0.45\linewidth]{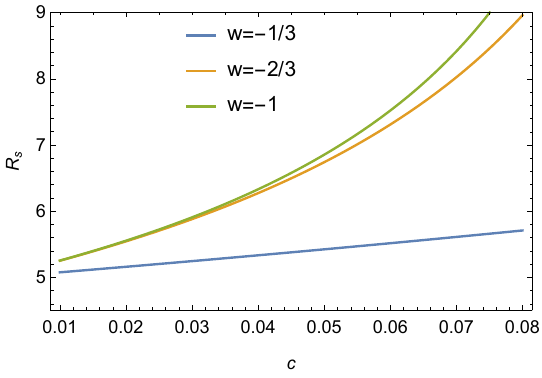}
    \caption{Variation  of  the  shadow  observable
$R_s$ of the EQG BH surrounded by
QF with the $c$ parameter for various state parameter $w$; here, $M=1$ and $\xi=1.5$. }
    \label{figa7}
\end{figure}

\begin{figure}[ht!]
\begin{center}
\includegraphics[scale=0.64]{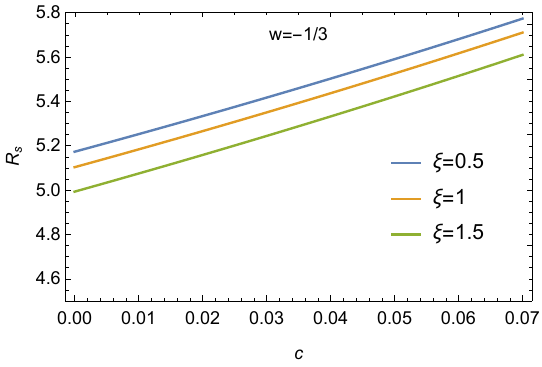}
\includegraphics[scale=0.64]{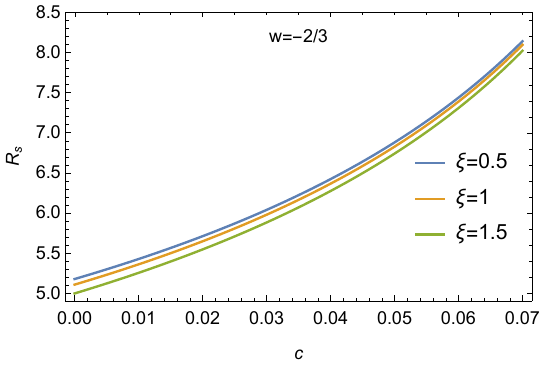}
\includegraphics[scale=0.64]{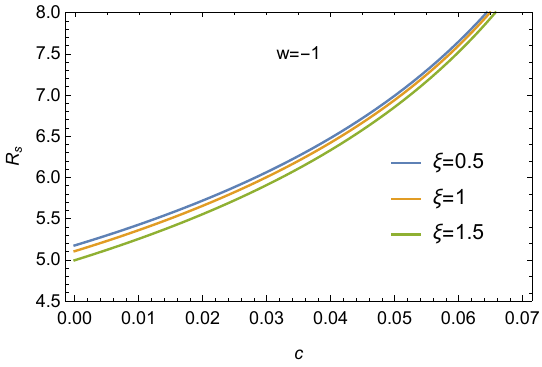}
\end{center}
\caption{Variation  of  the  shadow  observable
$R_s$ of the EQG BH surrounded by 
QF with the $c$ parameter for various quantum parameter $\xi$; here, $M=1$. }\label{figa2}
\end{figure}

In the limit where $c=0$, that is, without QF surrounding the BH spacetime, we recover the condition of  the EQG BH \cite{Yu-Heng}. Whereas if we choice for example $c=0.01$ we obtain 
\begin{equation}
    0\leq \xi \lesssim 2.547, \,\,  \mbox{for}\,\, w=-1/3, 
\end{equation}
\begin{equation}
    0\leq \xi \lesssim 3.530,\,\, \mbox{for}\, \, w=-2/3,
\end{equation}
\begin{equation}
   0\leq \xi \lesssim 3.633, \,\, \mbox{for}\,\, w=-1. 
\end{equation}
To obtain the range of permitted quantum-corrected parameters, obtained from the EHT observations of M87*, we generate Fig. \ref{figa8}. The figure shows that the allowable quantum corrected parameter range increases with the parameter $c$. 
We will now conduct a detailed investigation of the shadow radius to better understand the effect of the BH parameters on it. First, we plot the shadow radius using the $c$ parameter for various state parameters $w$ of the QF and a fixed quantum parameter $\xi=1.5$ (see Fig. \ref{figa7}). The figure indicates that when $w$ increases, so does the shadow radius. This suggests that the QF increases the shadow radius. Next, Fig. \ref{figa2} illustrates how $R_s$ is influenced by both parameters ($c$ and $\xi$). 

\begin{figure}[ht!]
    \centering
    \includegraphics[width=0.5\linewidth]{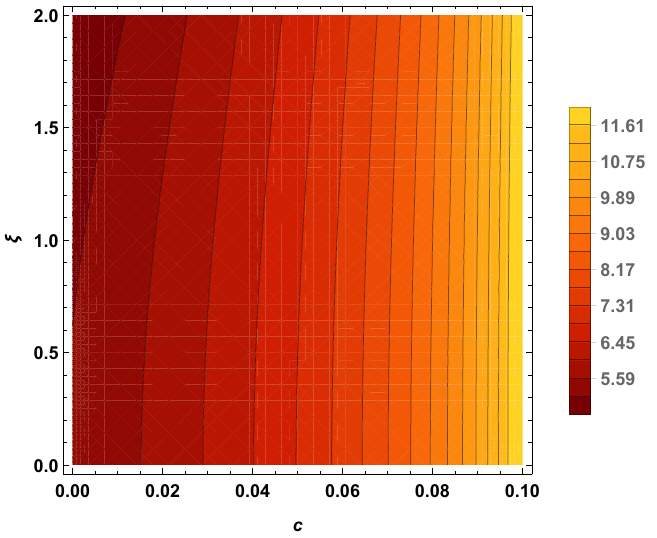}
    \caption{The contour maps of shadow observable
$R_s$ for the EQG BH with QF in the $(c,\xi)$ parameter space. Here, $M=1$ and $w=-2/3$.}
    \label{figa4}
\end{figure}

It can be seen from the Fig. \eqref{figa2} that, as parameter $c$ increases, the shadow radius increases, whereas quantum parameter $\xi$ decreases it.
Moreover,  to
show how the shadow observable
$R_s$ size varies with $(c,\xi)$ we plot the contours map for the EQG BH with QF in the $(c,\xi)$ parameter space in Fig. \ref{figa4}. It is evident that the shadow radius $R_s$ 
increases with the parameter $c$.

\section{Geodesics Analysis} \label{sec4}

In this section, we investigate the dynamics of both massless and massive test particles in the spacetime of a quantum-corrected/modified BH surrounded by a QF, as described by the metric (\ref{aa1}). Our focus is on analyzing the motion of these particles. Given that the spacetime (\ref{aa1}) is static and spherically symmetric, we can simplify the problem by restricting the analysis to geodesic motion in the equatorial plane, where $\theta=\pi/2$.

The Lagrangian density function in terms of the metric tensor $g_{\mu\nu}$, is expressed as $\mathcal{L} = \frac{1}{2}\,g_{\mu\nu}\,\dot{x}^{\mu}\,\dot{x}^{\nu}$ \cite{NPB, CJPHY, IJGMMP, AHEP1, AHEP2, AHEP3, AHEP4, AHEP5, EPJC}, where the dot represents differentiation with respect to the affine parameter $\tau$. By applying the metric (\ref{aa1}), we can explicitly write out this Lagrangian as follows:
\begin{equation}
    \mathcal{L}=\frac{1}{2}\,\left[-\mathcal{F}(r)\,\dot{t}^2+\frac{\dot{r}^2}{\mathcal{F}(r)}+r^2\,\dot{\phi}^2\right].\label{bb1}
\end{equation}

There are two constants of motions corresponding to the cyclic coordinates ($t, \phi$). These are respectively the energy ($\mathrm{E}$) and the conserved angular momentum ($\mathrm{L}$) and is given by
\begin{eqnarray}
    &&\mathrm{E}=\mathcal{F}(r)\,\dot{t},\label{bb2}\\
    &&\mathrm{L}=r^2\,\dot{\phi}.\label{bb3}
\end{eqnarray}

With these, geodesic equation for $r$ coordinate from Eq. (\ref{bb1}) can be obtained as:
\begin{equation}
    \dot{r}^2+V_\text{eff}=\mathrm{E}^2\label{bb4}
\end{equation}
which is equivalent to the one-dimensional equation of motion for a test particle of unit mass, with $\mathrm{E}^2$, where $V_\text{eff}$ represents the effective potential given by
\begin{equation}
    V_\text{eff}=\left(-\varepsilon+\frac{\mathrm{L}^2}{r^2}\right)\,\left[1-\frac{2\,M}{r}-\frac{c}{r^{3\,w+1}}+\frac{\xi^2}{r^2}\left( 1-\frac{2\,M}{r}-\frac{c}{r^{3\,w+1}} \right)^2 \right].\label{bb5}
\end{equation}
Here $\varepsilon=0$ for null geodesics and $-1$ for time-like.

From the expression in (\ref{bb5}), it is evident that the effective potential for geodesic motion is shaped by several factors. These include the quantum correction parameter $\xi$, the QF parameter $c$ associated with a specific state parameter $w$, the black hole mass $M$, and variations with the angular momentum $\mathrm{L}$. 

As an example, substituting the state parameter $w=-2/3$ into the Eq. (\ref{bb5}) results
\begin{equation}
    V_\text{eff}=\left(-\varepsilon+\mathrm{L}^2/r^2\right)\,\left\{1+\frac{\xi^2}{r^3}\,\left(r-2\,M-c\,r^2\right)\right\}\,\left(r-2\,M-c\,r^2\right).\label{bb5aa}
\end{equation}

In the limit where $c=0$, that is, without QF parameter, the effective potential from Eq. (\ref{bb5aa}) becomes,
\begin{equation}
    V_\text{eff}=\left(-\varepsilon+\mathrm{L}^2/r^2\right)\,\left(1+\frac{\xi^2}{r^3}\,(r-2\,M)\right)\,\left(r-2\,M\right)/r.\label{bb6}
\end{equation}
Moreover, in the limit where $\xi=0$, without any quantum correction but have QF, the effective potential from Eq. (\ref{bb5aa}) becomes 
\begin{equation}
    V_\text{eff}=\left(-\varepsilon+\mathrm{L}^2/r^2\right)\,\left(r-2\,M-c\,r^2\right)/r\label{bb7}
\end{equation}
which is similar to the result for the Kiselve BH solution with the state parameter $w=-2/3$.

Similarly, one can derive various expressions for the effective potential by choosing different values for the state parameter $w$. From the analysis of the effective potential for geodesic motion, it becomes clear that the quantum correction parameter $\xi$, along with the QF parameters $(c, w)$, collectively influence the effective potential for both null and time-like geodesics. Consequently, the effective potential is modified in comparison to the standard black hole solution.

\subsection{Null Geodesics: Motions of massless light-like particles} \label{sec4.1}

In this section, we focus on null geodesics to analyze photon trajectories and illustrate how the gravitational field of the selected black hole solution alters the photon’s path, causing it to deviate from its initial direction.

For photon particle, the effective potential for null geodesics ($\varepsilon=0$) from (\ref{bb5}) becomes 
\begin{equation}
    V_\text{eff}=\frac{\mathrm{L}^2}{r^2}\,\left[1+\frac{\xi^2}{r^2}\,\left( 1-\frac{2\,M}{r}-\frac{c}{r^{3\,w+1}} \right)\right]\,\left( 1-\frac{2\,M}{r}-\frac{c}{r^{3\,w+1}} \right).\label{cc1}
\end{equation}

\begin{figure}[ht!]
    \centering
    \includegraphics[width=0.4\linewidth]{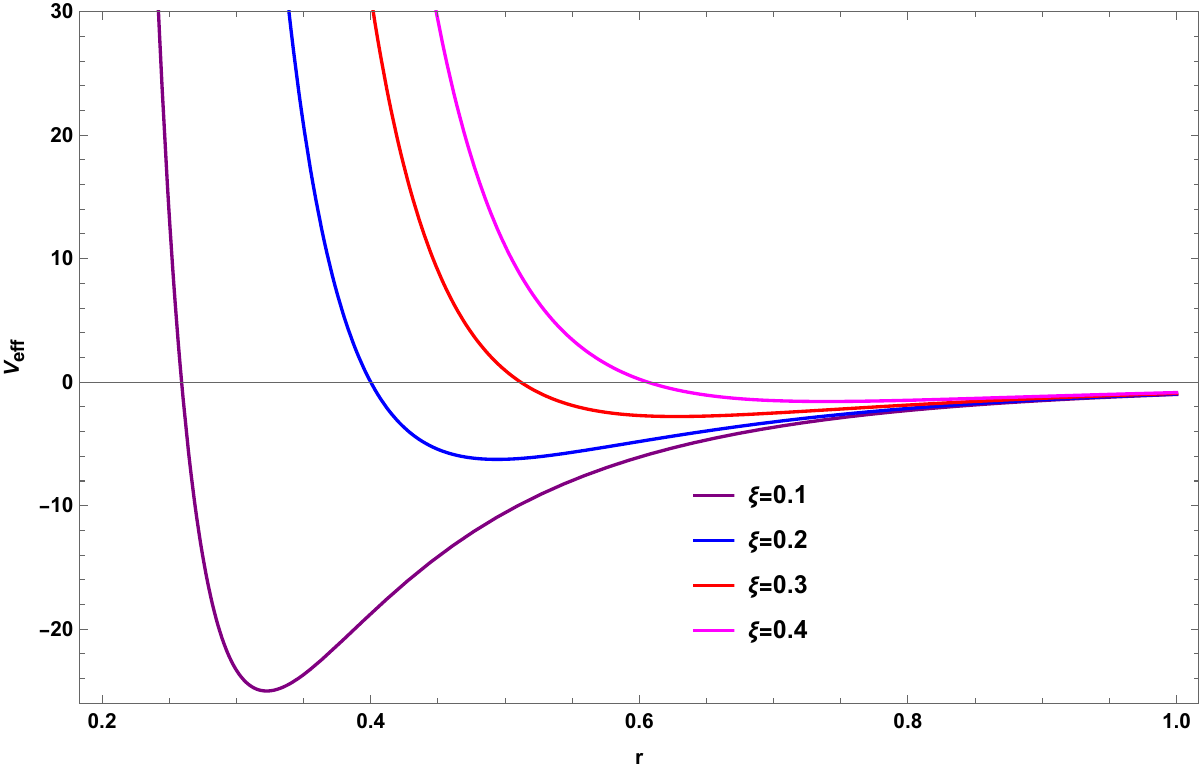}\quad\quad
    \includegraphics[width=0.4\linewidth]{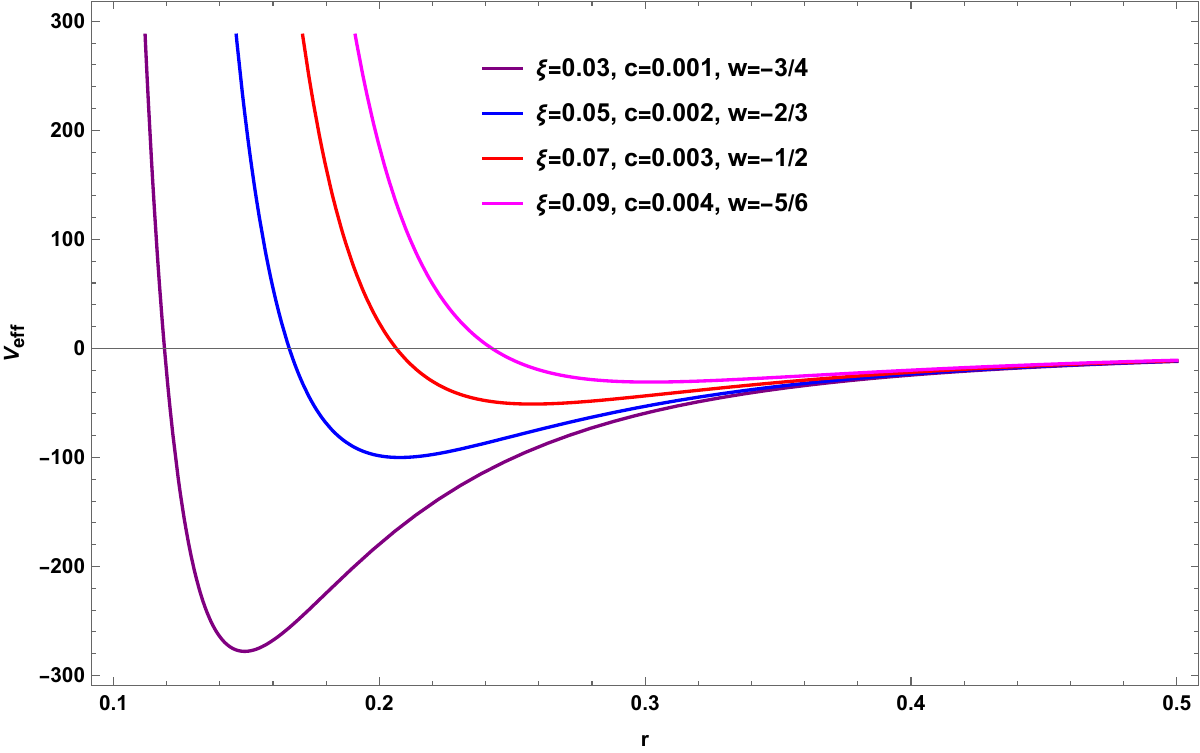}
    \caption{Illustration of null geodesics different values of quantum correction parameter $\xi$, the normalization constant $c$ with state parameter $w$. Left panel: $c=0.001$, $w=-2/3$. Here, we set $M=1=\mathrm{L}$.}
    \label{fig:null}
\end{figure}

In Fig. \ref{fig:null}, we depict the effective potential for null geodesics while varying the quantum correction parameter $\xi$, the normalization constant $c$ of the QF, and the state parameter $w$. In the left panel, we observe that as the value of the parameter $\xi$ increases, the effective potential also increases. In the right panel, an increase in the parameters $\xi$, $c$, and $w$ simultaneously results in an increase in the effective potential for null geodesics. 

For circular null geodesics, one have the conditions $\dot{r}=0$ and $\ddot{r}=0$ at a circular radius $r=r_c$. These conditions imply that $V_\text{eff}=\mathrm{E}^2$ and $V'_\text{eff}(r)=0$. Simplification of these conditions using Eq. (\ref{cc1}) for the state parameter $w=-2/3$ result the critical impact parameter:
\begin{eqnarray}
    \frac{1}{\beta_\text{critical}}=\frac{\mathrm{E}}{\mathrm{L}}=\frac{\sqrt{(r_c-2\,M-c\,r^2_c)\,\left[r^3_c+\xi^2\,(r_c-2\,M-c\,r^2_c)\right]}}{r^3_c}.\label{cc2}
\end{eqnarray}
And the radius of photon sphere relation
\begin{equation}
    2\,\mathcal{F}(r_c)=r_c\,\mathcal{F}'(r_c).\label{cc2a}
\end{equation}

To discuss stability of the circular orbits, one can determine a physical quantity called the Lyapunov exponent $\lambda_L$. For this, we find the following using the effective potential (\ref{cc1}) as:
\begin{equation}
    V''_\text{eff}=\frac{\mathrm{L}^2}{r^4}\,\left(r^2\,\mathcal{F}''(r)-4\,r\,\mathcal{F}'(r)+6\,\mathcal{F}(r)\right).\label{cc2b}
\end{equation}
At $r=r_c$ and using the relation (\ref{cc2a}), we find
\begin{equation}
    V''_\text{eff}(r_c)=\frac{\mathrm{L}^2}{r^4_{c}}\,\left(r^2_{c}\,\mathcal{F}''(r_c)-2\,\mathcal{F}(r_c)\right).\label{cc2c}
\end{equation}

The Lyapunov exponent for circular null orbits is therefore defined as:
\begin{equation}
    \lambda^\text{null}_L=\sqrt{-\frac{V''_\text{eff}(r_c)}{2\,\dot{t}^2}}=\sqrt{\mathcal{F}(r_c)\,\left(\frac{\mathcal{F}(r_c)}{r^2_{c}}-\frac{\mathcal{F}''(r_c)}{2}\right)},\label{lyapunov-null}
\end{equation}
where we have used $\dot{t}^2=\mathrm{L}^2/(r^2\,\mathcal{F})$.

Substituting the metric function $\mathcal{F}$ from (\ref{aa1}) and after simplification results (we have set state parameter $w=-2/3$)
\begin{equation}
    \lambda^\text{null}_L=\frac{1}{r^3}\sqrt{r^4-c\,r^5+[-36\, M^2 + 20\, M\, r - 2\, (1 + 4\, c\, M)\, r^2 + c^2\, r^4]\,\xi^2}.\label{lyapunov}
\end{equation}
From the above expression, it is clear that the Lyapunov exponent is influenced by factors, such as, the quantum correction parameter $\xi$, the QF constant parameter $c$, and the BH mass $M$. If the square of this exponent is negative, circular null orbits is stable otherwise chaotic.  



Now, we determine force on the photon particle using the effective potential given in Eq. (\ref{cc1}). The force can be defined in terms of the effective potential as, $\mathrm{F}=-V'_\text{eff}/2$. Thereby, using Eq. (\ref{cc1}), we find the force expression
\begin{equation}
    \mathrm{F}_\text{ph}=\frac{\mathrm{L}^2}{r^3}\,\left(1-\frac{3\,M}{r}-\frac{3\,c\,(w+1)/2}{r^{3\,w+1}}\right)\left[1+\frac{2\,\xi^2}{r^2}\,\left( 1-\frac{2\,M}{r}-\frac{c}{r^{3\,w+1}} \right)\right].\label{cc3}
\end{equation}
From the above expression (\ref{cc3}), it is clear that the force on photon particle is influenced by the quantum correction parameter $\xi$, the QF parameters $(w,c)$, and the BH mass $M$.

For a specific state parameter, $w=-2/3$, the force from Eq. (\ref{cc3}) reduces as: 
\begin{equation}
    \mathrm{F}_\text{ph}=\frac{\mathrm{L}^2}{r^3}\,\left(1-\frac{3\,M}{r}-\frac{c\,r}{2}\right)\left[1+\frac{2\,\xi^2}{r^2}\,\left( 1-\frac{2\,M}{r}-c\,r \right)\right].\label{cc4}
\end{equation}

In the limit where $c=0$, without QF, from Eq. (\ref{cc4}) we find
\begin{equation}
    \mathrm{F}_\text{ph}=\frac{\mathrm{L}^2\,(r-3\,M)\,\left(r^3+2\,\xi^2\,(r-2\,M)\right)}{r^7}.\label{cc3b}
\end{equation}

Thereby, comparing Eqs. (\ref{cc4}) and (\ref{cc3b}), it is evident that the presence of quintessence field parameters $(w, c)$ in selected BH solution affects the force on photon particle. Moreover, 
 the presence of quantum correction parameter $\xi$ alters the force in comparison to the Kiselve BH solution for a specific state parameter $w=-2/3$.

\begin{figure}[ht!]
    \centering
    \includegraphics[width=0.45\linewidth]{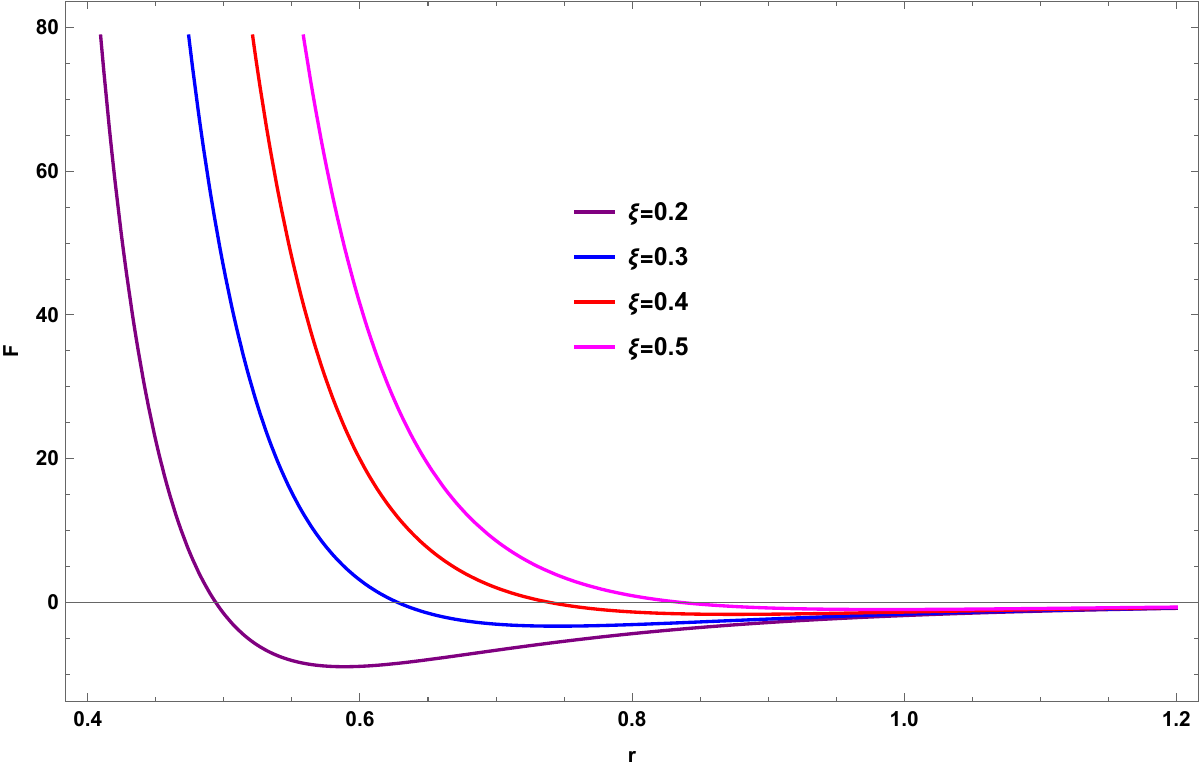}\quad\quad
    \includegraphics[width=0.45\linewidth]{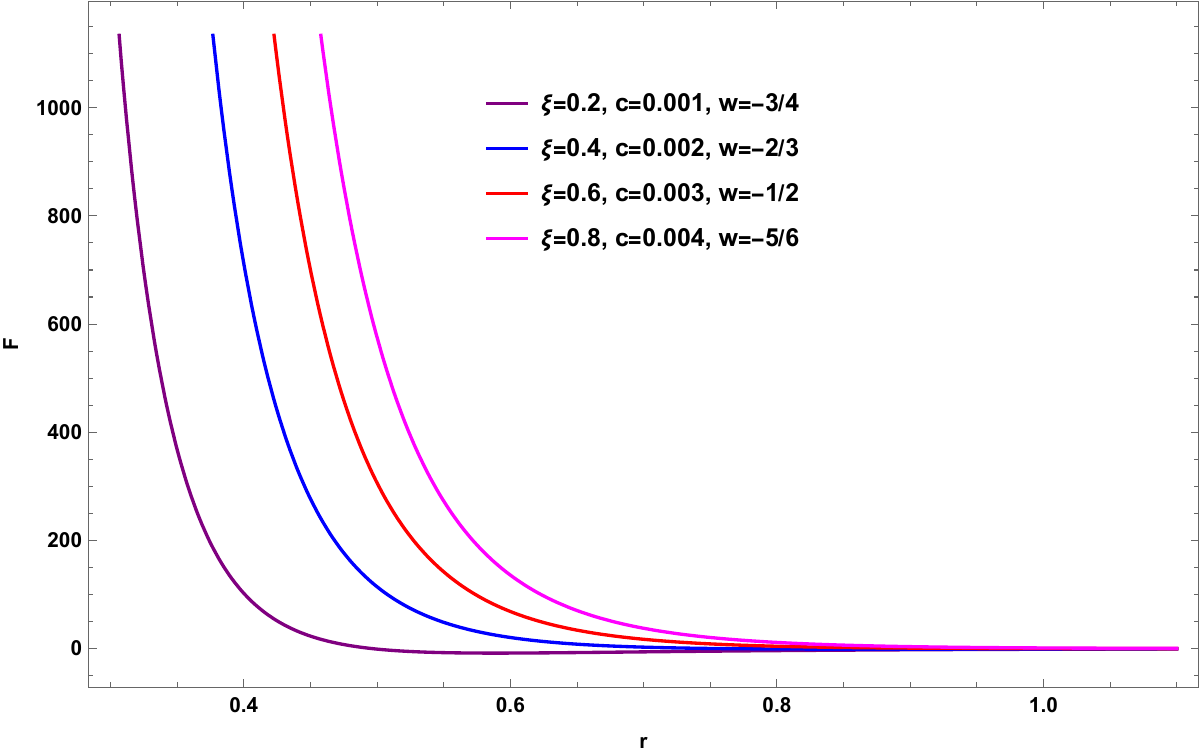}
    \caption{Illustration of force on photon light for different values of quantum correction parameter $\xi$, the normalization constant $c$ with state parameter $w$. Left panel: $c=0.001$, $w=-2/3$. Here, we set $M=1=\mathrm{L}$.}
    \label{fig:force}
\end{figure}

In Fig. \ref{fig:force}, we present the force acting on massless photon particles while varying the quantum correction parameter $\xi$, the normalization constant $c$ of the QF, and the state parameter $w$. In the left panel, we observe that as the quantum correction parameter $\xi$ increases, the force experienced by the photon particles also increases. This suggests that the quantum correction contributes to a stronger interaction between the photon and the gravitational field. In the right panel, we notice that an increase in the parameters $\xi$, $c$, and $w$ leads to a combined effect that further enhances the force acting on the photon particles. This highlights the role of both the quantum correction and the properties of the QF in influencing the photon dynamics. Overall, the figure demonstrates how the gravitational field of a quantum-corrected BH, surrounded by a QF, affects the propagation of photon light as it passes near the BH, revealing the complex interplay between quantum effects and cosmic structures.

Finally, we derive the photon ray trajectory equation in the gravitational field of a BH that has been quantum-corrected and is surrounded by QF.

Using Eqs. (\ref{bb3}), (\ref{bb4}) and (\ref{cc1}), we define the following quantity
\begin{equation}
    \frac{\dot{r}^2}{\dot{\phi}^2}=\left(\frac{dr}{d\phi}\right)^2=r^4\,\left[\frac{1}{\beta^2}-\frac{1}{r^2}\,\left\{1-\frac{2\,M}{r}-\frac{c}{r^{3\,w+1}}+\frac{\xi^2}{r^2}\left( 1-\frac{2\,M}{r}-\frac{c}{r^{3\,w+1}} \right)^2 \right\}\right].\label{cc5}
\end{equation}
where $\beta=\mathrm{L}/\mathrm{E}$ is the impact parameter for photon light.

Transforming to a new variable via $u=\frac{1}{r}$ into the Eq. (\ref{cc5}]) results the following second-order differential equation given by:
\begin{equation}
    \left(\frac{du}{d\phi}\right)^2+u^2=\frac{1}{\beta^2}+2\,M\,u^3+c\,u^{3\,w+3}-\xi^2\,u^4\,\Big(1-2\,M\,u-c\,u^{3\,w+1} \Big)^2.\label{cc6}
\end{equation}
Equation (\ref{cc6}) represents the photon ray trajectory equation in the gravitational field of a BH, described by an effective quantum-corrected solution, with the influence of QF.

In the limit where $\xi=0$, that is, with no quantum corrections into the BH solution, from Eq. (\ref{cc6}) we find the following second-order differential equation given by:
\begin{equation}
    \left(\frac{du}{d\phi}\right)^2+u^2=\frac{1}{\beta^2}+2\,M\,u^3+c\,u^{3\,w+3}\label{cc7}
\end{equation}
which is similar to the photon trajectory equation in Kiselve BH solution.

Moreover, in the limit where $c=0$, that is, without  QF, from Eq. (\ref{cc6}) we find the following second-order differential equation given by: 
\begin{equation}
    \left(\frac{du}{d\phi}\right)^2+u^2=\frac{1}{\beta^2}+2\,M\,u^3-\xi^2\,u^4\,\Big(1-2\,M\,u\Big)^2.\label{cc8}
\end{equation}
Equation (\ref{cc8}) is the photon light trajectory equation in an effective quantum corrected BH solution without QF.

Through a comparison of Eqs. (\ref{cc6}), (\ref{cc7}), and (\ref{cc8}), it becomes evident that the photon trajectory equation is altered by the dual influence of the quantum correction and the QF, collectively shaping the gravitational field produced by the space-time geometry specified in (\ref{aa1}).

\subsection{Time-like Geodesics: Motions of time-like particles} \label{sec4.2}

In this part, we study dynamics of time-like particles in the gravitational field produced by the space-time geometry, specified in (\ref{aa1}).

For massive test particles, $\varepsilon=-1$, and hence, the effective potential for time-like particles from (\ref{bb5}) becomes
\begin{equation}
    V_\text{eff}=\left(1+\frac{\mathrm{L}^2}{r^2}\right)\,\left[1-\frac{2\,M}{r}-\frac{c}{r^{3\,w+1}}+\frac{\xi^2}{r^2}\left( 1-\frac{2\,M}{r}-\frac{c}{r^{3\,w+1}} \right)^2 \right].\label{dd1}
\end{equation}

\begin{figure}[ht!]
    \centering
    \includegraphics[width=0.4\linewidth]{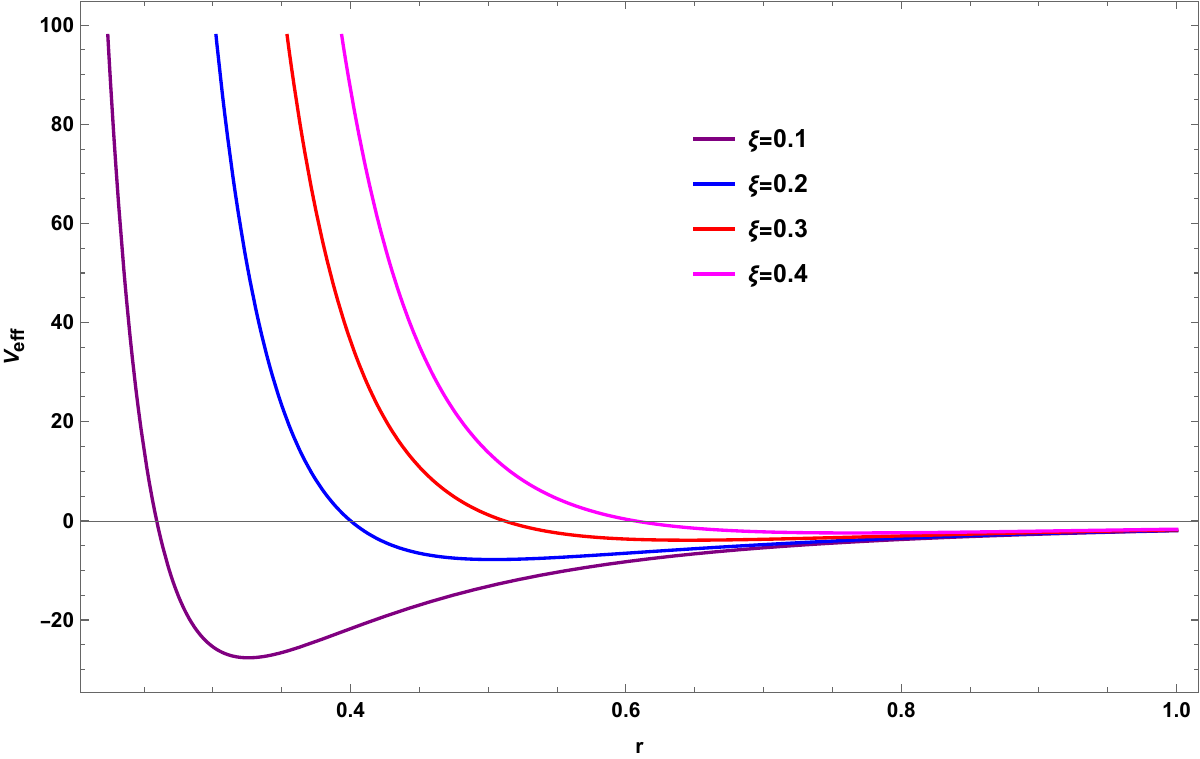}\quad\quad
    \includegraphics[width=0.4\linewidth]{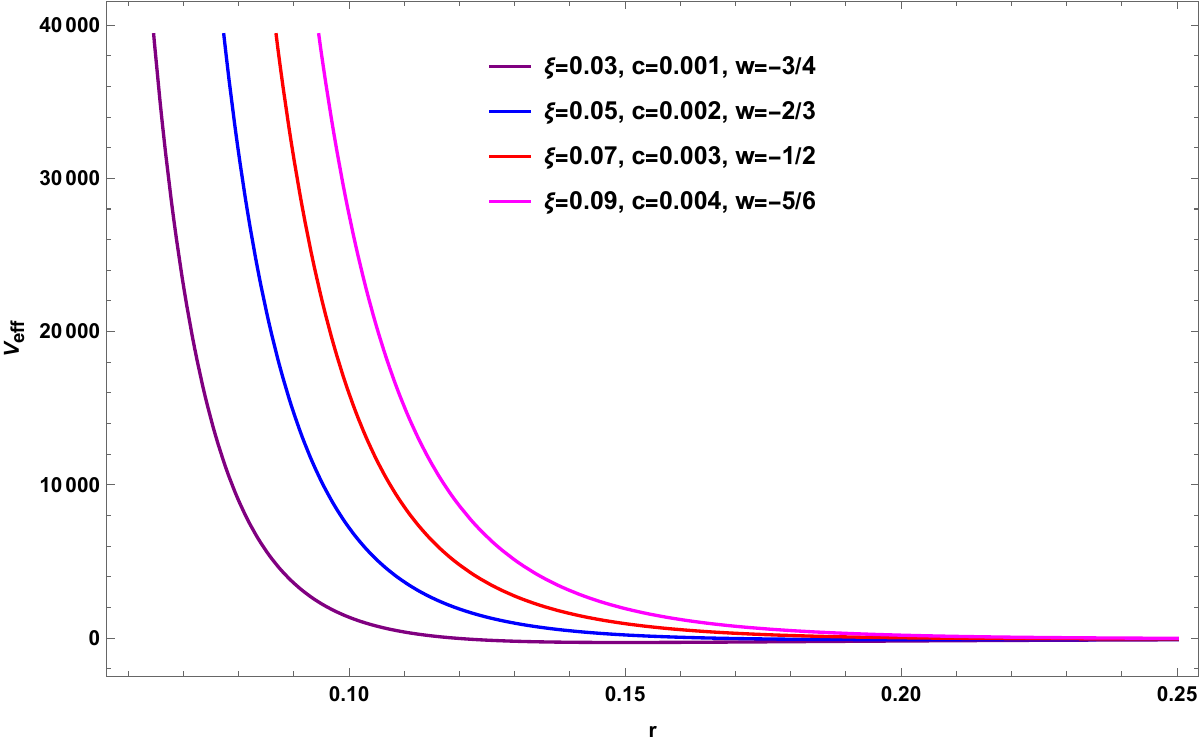}
    \caption{Illustration of time-like geodesics different values of quantum correction parameter $\xi$, the normalization constant $c$ with state parameter $w$. Left panel: $c=0.001$, $w=-2/3$. Here, we set $M=1=\mathrm{L}$.}
    \label{fig:time-like}
\end{figure}

In Fig. \ref{fig:time-like}, we depict the effective potential for time-like geodesics while varying the quantum correction parameter $\xi$, the normalization constant $c$ of the QF, and the state parameter $w$. In the left panel, we observe that as the value of the parameter $\xi$ increases, the effective potential also increases. In the right panel, an increase in the parameters $\xi$, $c$, and $w$ simultaneously results in an increase in the effective potential for null geodesics. Overall, the figure demonstrates how the gravitational field of a quantum-corrected BH, surrounded by a QF, affects the effective potential for time-like particles as it passes near the BH, showing the complex interplay between quantum effects and cosmic structures.

For circular motions of time-like particles in the equatorial plane, the conditions $\dot{r}=0$ and $\ddot{r}=0$ lead to the relations $\mathrm{E}^2=V_\text{eff}(r)$ and $V'_\text{eff}(r)=0$, respectively, where $V_\text{eff}$ is defined in Eq. (\ref{dd1}). Simplifying these relations gives the expression for the angular momentum, as follows:
\begin{equation}
    \mathrm{L}=r\,\sqrt{\frac{\frac{M}{r}+\frac{c\,(3\,w+1)/2}{r^{3\,w+1}}+\frac{\xi^2}{r^2}\,\left( 1-\frac{2\,M}{r}-\frac{c}{r^{3\,w+1}} \right)\,\left(\frac{4\,M}{r}-1+\frac{c\,(3\,w+2)}{r^{3\,w+1}} \right)}{\left(1-\frac{3\,M}{r}-\frac{3\,c\,(w+1)/2}{r^{3\,w+1}}\right)\left[1+\frac{2\,\xi^2}{r^2}\,\left( 1-\frac{2\,M}{r}-\frac{c}{r^{3\,w+1}} \right)\right]}}.\label{dd3}
\end{equation}
And the particles' energy as,  
\begin{equation}
    \mathrm{E}_{\pm}=\pm\,\frac{1-\frac{2\,M}{r}-\frac{c}{r^{3\,w+1}}+\frac{\xi^2}{r^2}\left( 1-\frac{2\,M}{r}-\frac{c}{r^{3\,w+1}} \right)^2}{\sqrt{\left(1-\frac{3\,M}{r}-\frac{3\,c\,(w+1)/2}{r^{3\,w+1}}\right)\left[1+\frac{2\,\xi^2}{r^2}\,\left( 1-\frac{2\,M}{r}-\frac{c}{r^{3\,w+1}} \right)\right]}}.\label{dd4}
\end{equation}

From the above expression for the energy of time-like particles, we observe that the energy approaches $\mathrm{E}_{\pm} \to \pm\,1$ as $r \to \infty$. Thus, the maximum energy of time-like particles is unity (1). 

\begin{figure}[ht!]
    \centering
    \includegraphics[width=0.45\linewidth]{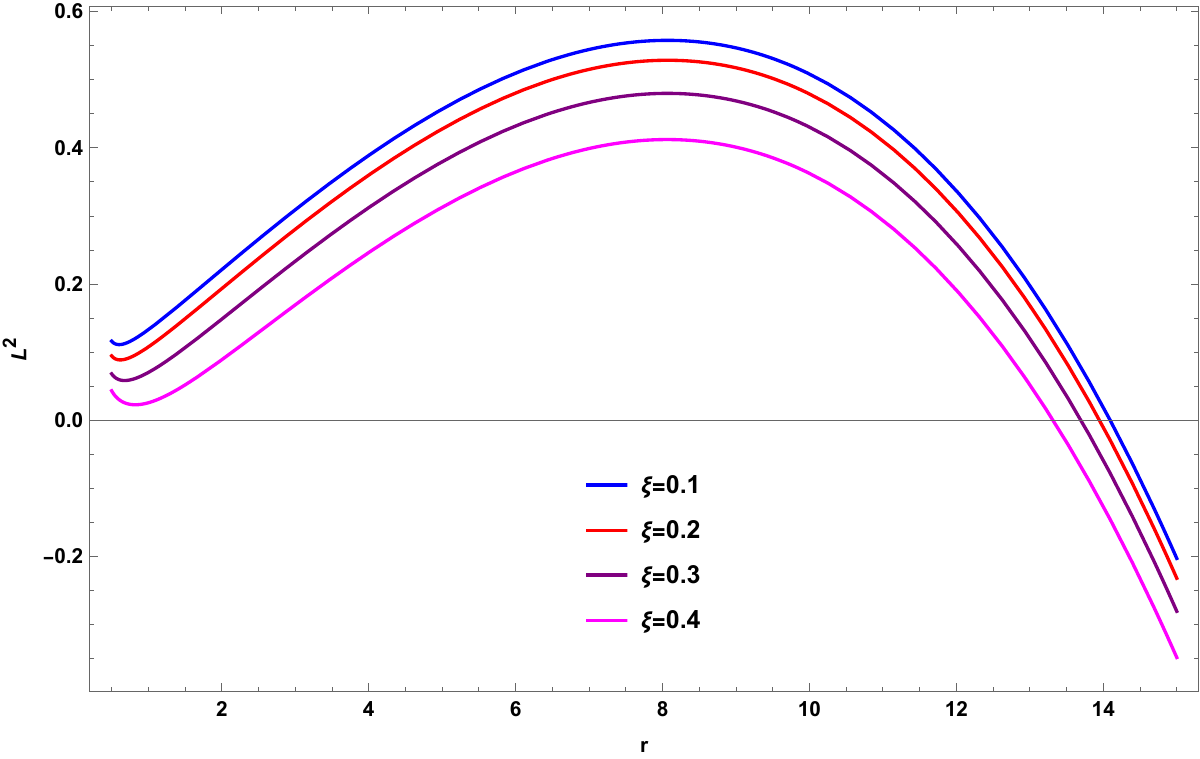}\quad\quad
    \includegraphics[width=0.45\linewidth]{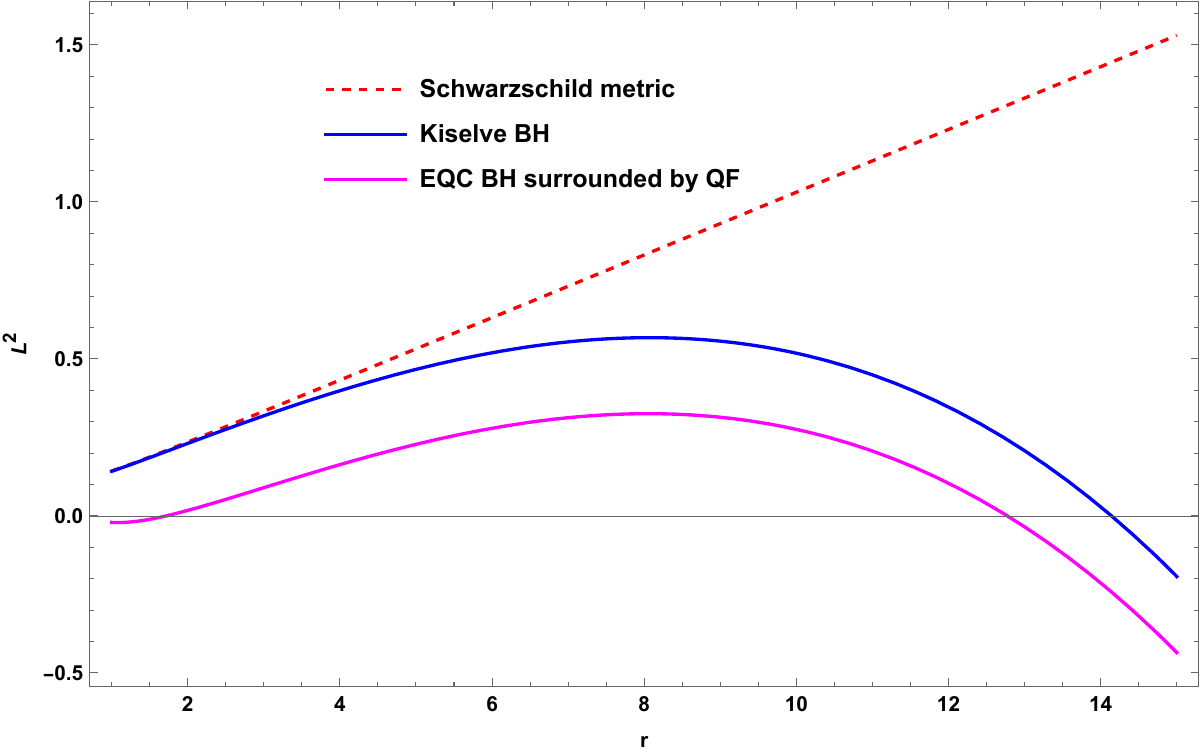}
    \caption{Illustration of angular momentum for time-like particles in circular orbits in the equatorial plane varying quantum correction parameter $\xi$. Left panel: $M=0.1$, $c=0.001$, $w=-2/3$, right panel: $M=0.1$, $c=0.001$, $w=-2/3$ and $\xi=0.5$.}
    \label{fig:angular-momentum}
\end{figure}

In Fig. \ref{fig:angular-momentum}, we present the angular momentum of time-like particles in circular orbits in the equatorial plane, varying the quantum correction parameter $\xi$, while keeping the normalization constant $c = 0.001$ of the QF and the state parameter $w = -2/3$ fixed. In the left panel, we observe that as the value of the quantum correction parameter $\xi$ increases, the square of the angular momentum decreases. In the right panel, a comparison of the angular momentum is shown for different BHs, illustrating how the angular momentum changes across varying BH configurations.

\begin{figure}[ht!]
    \centering
    \includegraphics[width=0.45\linewidth]{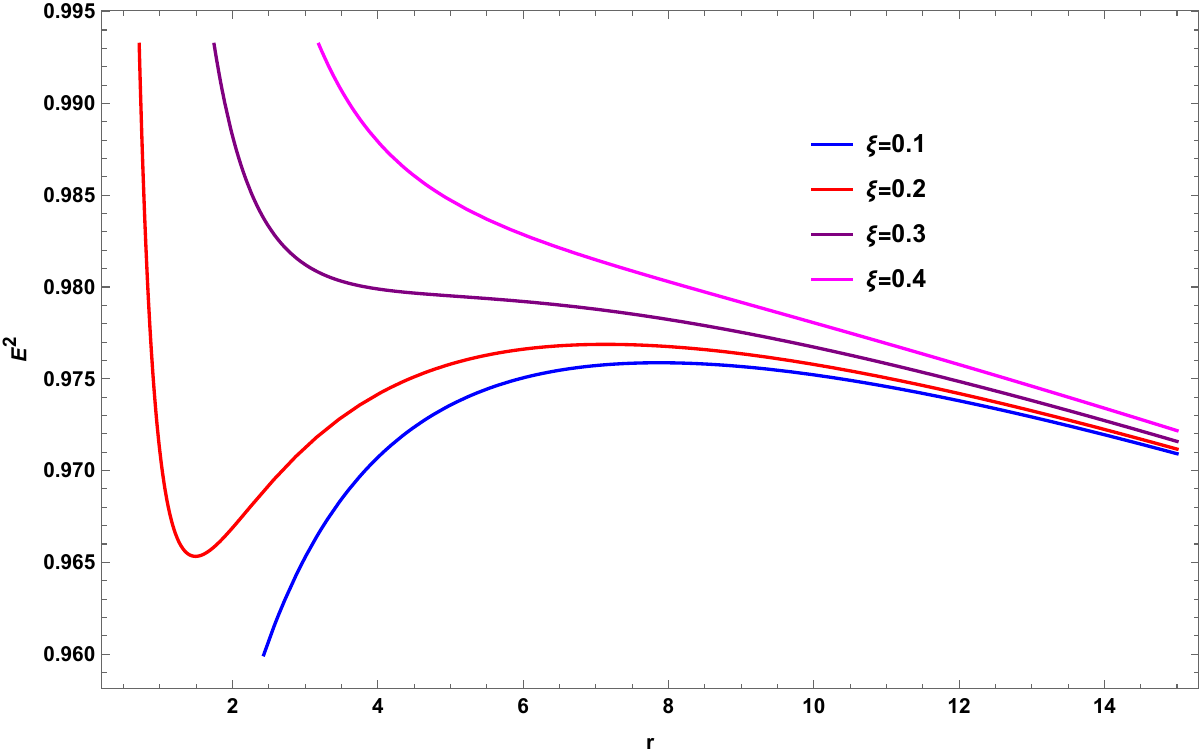}\quad\quad
    \includegraphics[width=0.45\linewidth]{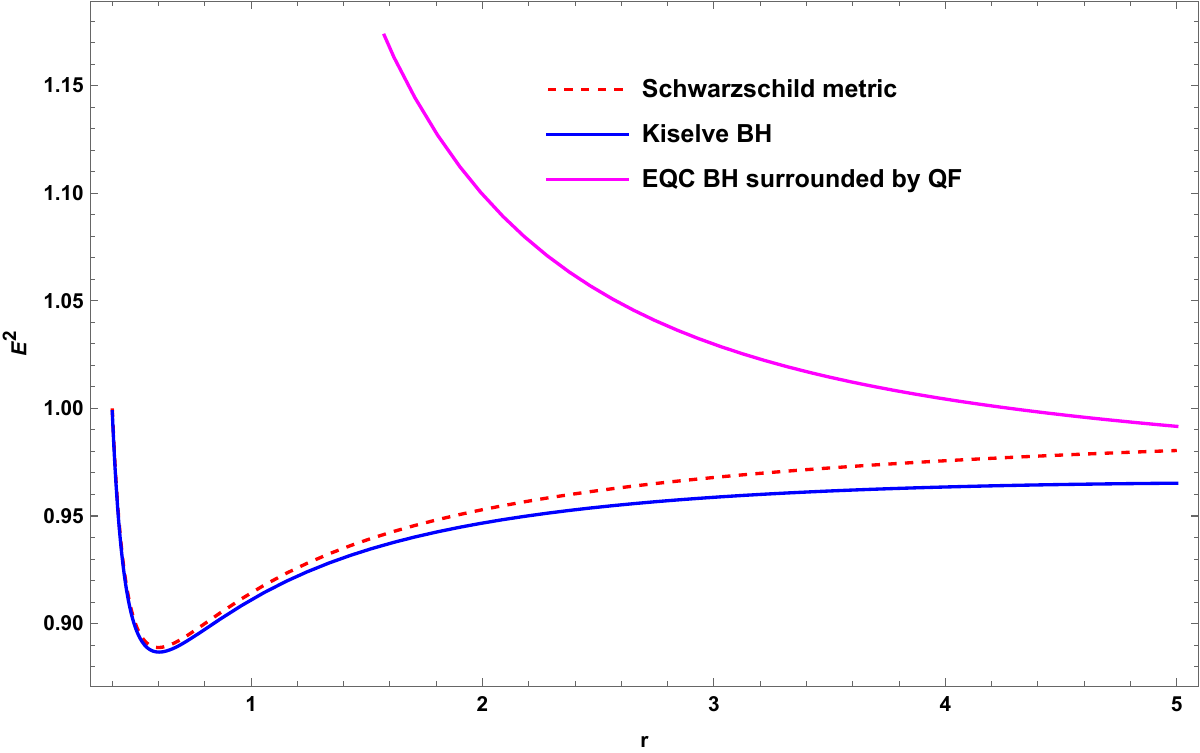}
    \caption{Illustration of energy for time-like particles in circular orbits in the equatorial plane varying quantum correction parameter $\xi$. Left panel: $M=0.1$, $c=0.001$, $w=-2/3$, right panel: $M=0.1$, $c=0.001$, $w=-2/3$ and $\xi=0.5$.}
    \label{fig:energy}
\end{figure}

In Fig. \ref{fig:energy}, we illustrate the energy of time-like particles in circular orbits in the equatorial plane, varying the quantum correction parameter $\xi$, while keeping the normalization constant $c = 0.001$ of the QF and the state parameter $w = -2/3$ fixed. In the left panel, we see that as the value of the quantum correction parameter $\xi$ increases, the square of the energy also increases. In the right panel, a comparison of the energy is shown for different BHs, illustrating how the energy of time-like particles changes across varying BH configurations.

In the limit where $\xi=0$, with no quantum correction, these physical quantities ($\mathrm{L}$,$\mathrm{E}_{\pm}$) for time-like particles reduce to the results as follows for the Kiselev BH solution given by
\begin{eqnarray}
    \mathrm{L}=r\,\sqrt{\frac{\frac{M}{r}+\frac{c\,(3\,w+1)/2}{r^{3\,w+1}}}{\left(1-\frac{3\,M}{r}-\frac{3\,c\,(w+1)/2}{r^{3\,w+1}}\right)}},\quad\quad
    \mathrm{E}_{\pm}=\pm\,\frac{1-\frac{2\,M}{r}-\frac{c}{r^{3\,w+1}}}{\sqrt{\left(1-\frac{3\,M}{r}-\frac{3\,c\,(w+1)/2}{r^{3\,w+1}}\right)}}.\label{dd4aa}
\end{eqnarray}

Moreover, in the limit where $c=0$, with no QF, these quantities ($\mathrm{L},\mathrm{E}$) reduces as,
\begin{eqnarray}
    \mathrm{L}=r\,\sqrt{\frac{\frac{M}{r}+\frac{\xi^2}{r^2}\,\left( 1-\frac{2\,M}{r} \right)\,\left(\frac{4\,M}{r}-1\right)}{\left(1-\frac{3\,M}{r}\right)\left[1+\frac{2\,\xi^2}{r^2}\,\left( 1-\frac{2\,M}{r}\right)\right]}},\quad\quad
    \mathrm{E}_{\pm}=\pm\,\frac{1-\frac{2\,M}{r}+\frac{\xi^2}{r^2}\left( 1-\frac{2\,M}{r}\right)^2}{\sqrt{\left(1-\frac{3\,M}{r}\right)\left[1+\frac{2\,\xi^2}{r^2}\,\left( 1-\frac{2\,M}{r}\right)\right]}}.\label{dd4aaa}
\end{eqnarray}

By comparing Eqs. (\ref{dd3}), (\ref{dd4}), (\ref{dd4aa}), and (\ref{dd4aaa}), it is evident that both the angular momentum and energy of time-like particles on circular orbits in the equatorial plane are modified by the combined effects of the quantum correction and the QF, which together influence the gravitational field produced by the space-time geometry described in (\ref{aa1}).

Moreover, we determine angular velocity for time time-like test particles on circular orbits in the equatorial plane which is given by
\begin{equation}
    \Omega=\sqrt{\frac{\mathcal{F}'(r)}{2\,r}}=\frac{1}{r}\,\sqrt{\frac{M}{r}+\frac{c\,(3\,w+1)/2}{r^{3\,w+1}}+\frac{\xi^2}{r^2}\,\left(1-\frac{2\,M}{r}-\frac{c}{r^{3\,w+1}} \right)\,\left(\frac{4\,M}{r}-1+\frac{c\,(3\,w+2)}{r^{3\,w+1}} \right)}.\label{dd5}
\end{equation}

In the limit where $c=0$, with no QF, this angular velocity for time-like particles from Eq. (\ref{dd5}) reduces as,
\begin{equation}
    \Omega=\frac{1}{r}\,\sqrt{\frac{M}{r}-\frac{\xi^2}{r^2}\,\left(1-\frac{2\,M}{r}\right)\,\left(1-\frac{4\,M}{r}\right)}.\label{dd5aa}
\end{equation}

By comparing Eqs. (\ref{dd5}) and (\ref{dd5aa}), it is evident that  the angular velocity for time-like particles on circular orbits in the equatorial plane is modified by the QF for a particular state parameter, and thus, shifts the result.

\begin{figure}[ht!]
    \centering
    \includegraphics[width=0.5\linewidth]{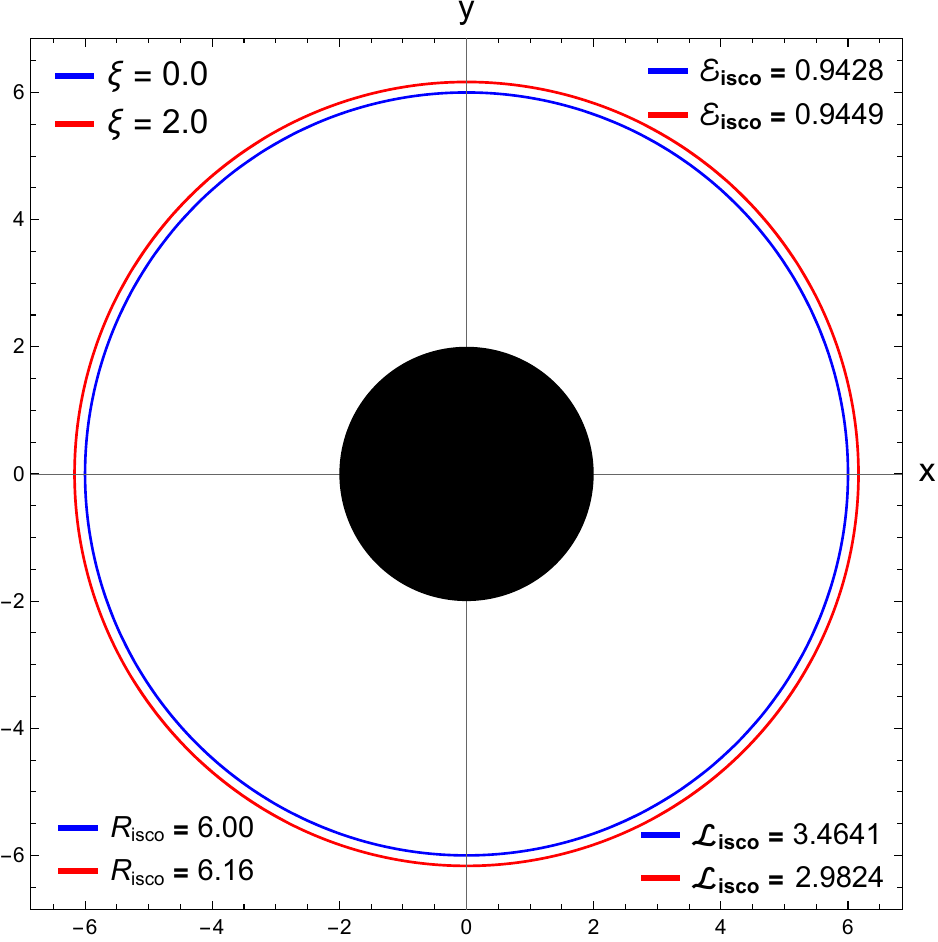}
    \caption{Illustration of the ISCO for different values of the quantum correction parameter $\xi$. To isolate the pure effect of the quantum correction parameter $\xi$ on $R_{\text{isco}}$, we plot $R_{\text{isco}}$ while assuming the QF parameter $c = 0$. Here, we set $M = 1$.}
    \label{fig:isco}
\end{figure}

In Fig.~\ref{fig:isco}, we illustrate the effect of the quantum correction parameter $\xi$ on the innermost stable circular orbit radius $R_{\text{isco}}$, the specific energy $\mathcal{E}_{\text{isco}}$, and the specific angular momentum $\mathcal{L}_{\text{isco}}$. From Fig.~\ref{fig:isco}, it is evident that as $\xi$ increases, both $R_{\text{isco}}$ and $\mathcal{E}_{\text{isco}}$ increase, while $\mathcal{L}_{\text{isco}}$ decreases. The increase in $R_{\text{isco}}$ suggests that the gravitational influence also strengthens with increasing $\xi$. However, the effect of the quantum correction parameter $\xi$ on $R_{\text{isco}}$ is minimal.

Using the same method applied to derive Eq.~\eqref{cc6}, we obtain the following trajectory equation for a massive particle in the case of $w=-2/3$
\begin{equation} \label{eqom}
    \left(\frac{du}{d\phi}\right)^2=\frac{\mathcal{E}^2}{\mathcal{L}^2}+\frac{ \left(c-u (1-2 M u)-u (c-u (1-2 M u))^2\xi ^2\right)\left(1+u^2\mathcal{L}^2 \right)}{  \mathcal{L}^2 u}\, ,
\end{equation}
where $\mathcal{E}$ and $\mathcal{L}$ are the specific energy and the specific angular momentum, respectively. By differentiating Eq.~\eqref{eqom} with respect to $\phi$, we derive the geodesic equation governing the motion of a massive test particle in the vicinity of a quantum corrected BH surrounded by QF
\begin{figure}[ht!]
    \centering
    \includegraphics[width=0.45\linewidth]{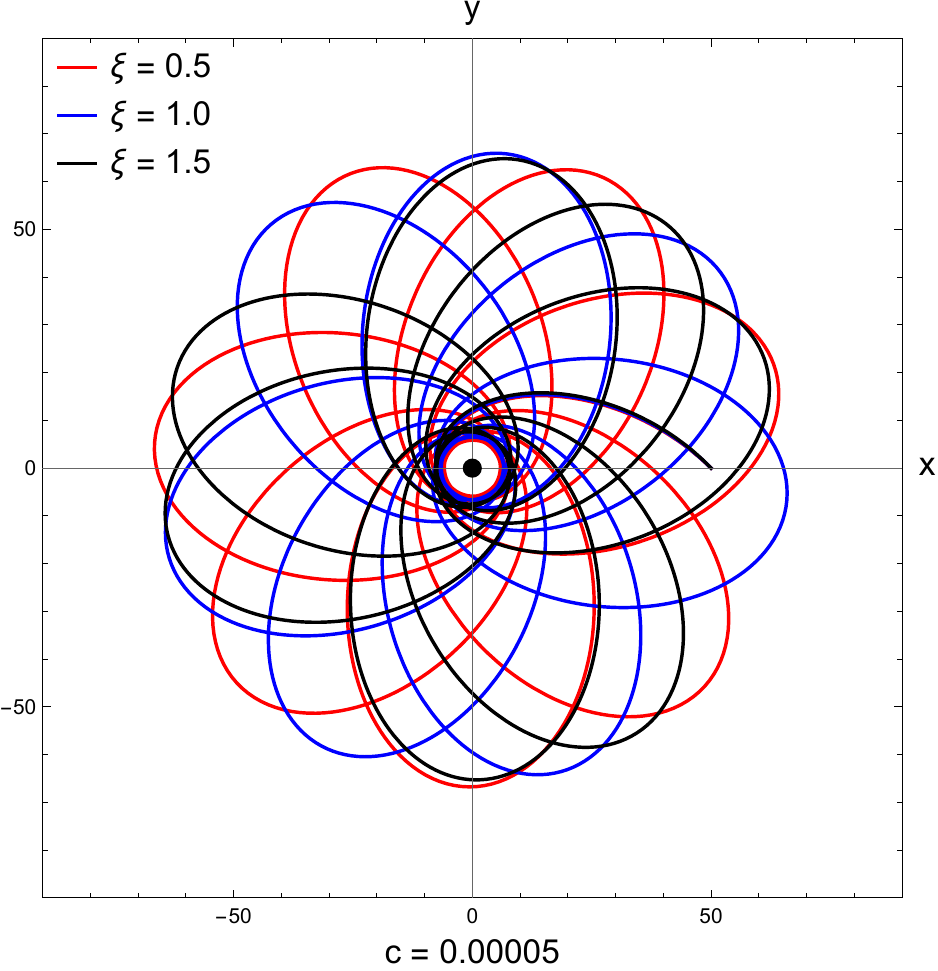}\quad\quad
    \includegraphics[width=0.45\linewidth]{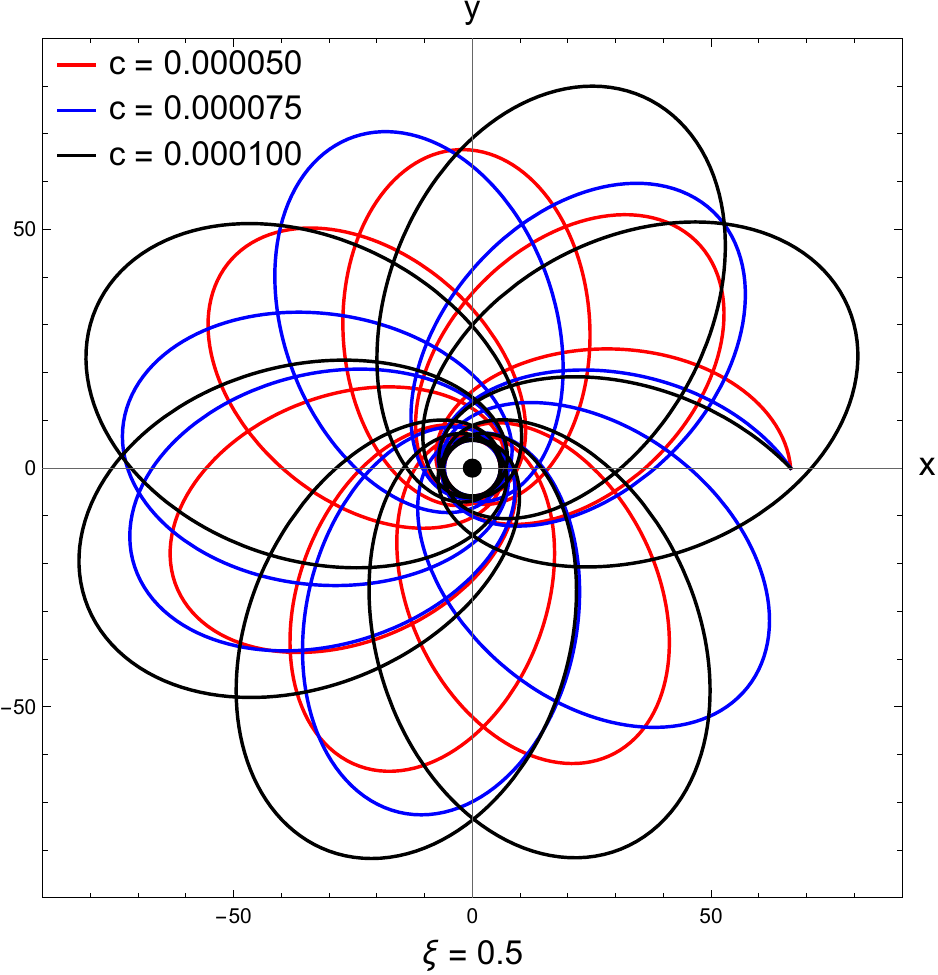}
    \caption{Illustration of the trajectories of a massive particle moving in the equatorial plane ($z=0$) around the quantum corrected BH surrounded by QF.  The trajectories are analyzed for different values of the quantum correction parameter $\xi$ and the QF parameter $c$. 
    Left panel: The influence of $\xi$ on the orbital motion of a particle with initial inverse radial coordinate $u = 1/r = 0.02$. 
    Right panel: The impact of $c$ on the trajectory of a particle with initial inverse radial coordinate $u = 1/r = 0.015$. 
    In both the left and right panels, the BH mass is set to $M=1$, while the specific energy and specific angular momentum of the particle are fixed at $\mathcal{E} = 0.985$ and $\mathcal{L} = 4$, respectively.
    }
    \label{fig:geodesics}
\end{figure}
\begin{align} \label{orbital-equation}
    \frac{d^2 u}{d\phi^2} =c+\frac{M}{\mathcal{L}^2}- u+3 M u^2-\frac{c}{2\mathcal{L}^2 u^2}-\frac{\xi ^2 (c-u (1-2 M u)) \left(4 M u-1+\mathcal{L}^2 u (c-2 u (1-3 M u))\right)}{\mathcal{L}^2} \, .
\end{align}
Fig.~\ref{fig:geodesics} illustrates the effect of the quantum correction parameter $\xi$ and the quintessential parameter $c$ on the trajectory of a massive particle. In the left panel of Fig.~\ref{fig:geodesics}, the geodesic trajectory is plotted for different values of $\xi$ under the same initial conditions ($u = 1/r = 0.02$, $\mathcal{E} = 0.985$, $\mathcal{L} = 4$). We observe that the effect of $\xi$ on the size of the orbit is minimal, as mentioned previously in Fig.~\ref{fig:isco}. This effect is evident from the curves of the red, blue, and black rings closest to the BH, formed by the trajectory and ordered according to increasing $\xi$. This occurs because, as noted earlier, $R_{\text{isco}}$ also increases with increasing $\xi$. Furthermore, although the initial conditions remain the same, an increase in $\xi$ causes the trajectories to deviate differently from one another. The right panel of Fig.~\ref{fig:geodesics} exhibits the effect of the quintessential parameter $c$ on the trajectory of time-like particles around the BH. It is observed that as $c$ increases, the size of the orbit expands. This suggests that $c$ has a repulsive nature, effectively reducing the gravitational attraction.

Finally, we investigate the stability of circular orbits for time-like particles using the Lyapunov exponent. This exponent is defined as:
\begin{equation}
    \lambda_{L}=\sqrt{-\frac{V'_\text{eff}}{2\,\dot{t}^2}},\quad \dot{t}=\frac{\mathrm{E}}{\mathcal{F}(r)}=\sqrt{\frac{2}{2\,\mathcal{F}(r)-r\,\mathcal{F}'(r)}}.\label{dd6}
\end{equation}

Using Eq. (\ref{dd1}), we find the following expression of the Lyapunov exponent given as,
\begin{eqnarray}
    \lambda^2_{L}&=&\frac{1}{2}\,r^{-2\,(5 + 6\, w)}\,\Bigg[12\, c^4\, (1 + w)\, (2 + 3\, w)\, \xi^4 -c^3\, r^{3\, w}\,\xi^2\,\Big\{9\,r^3\,(1 + w)\,(1 + 3\,w)+2\, \Big\{37\,r+3\,(8+3\,w)\times\nonumber\\
      &&\Big(3\,r\,w-2\,M\,(2+3\,w)\Big)\Big\}\,\xi^2\Big\}-2\, r^{12\,w}\, \Big\{M\, r^6\, (-6\, M + r) + 
      3\, M\, r^3\, \left(12\, M^2 - 8\, M\, r + r^2\right) \xi^2\nonumber\\
      &&-2\, (-2\, M + r)^2\, \left(24\, M^2 - 13\, M\, r + 2\, r^2\right)\,\xi^4\Big\}+3\, c^2\, r^{6\,w}\,\Big\{r^6\,(1 + w)\, (1 + 3\, w)\nonumber\\
      &&+2\, r^3\, \Big\{2\, r + 6\, r\, w - 3\, M\, (3 + 8\, w)\Big\}\,\xi^2
      +4\,\Big\{12\, M^2\, (4 + 5\, w) + r^2\, (7 + 9\, w)-M\, r\, (37 + 48\, w)\Big\}\,\xi^4\Big\}\nonumber\\
      &&+c\,r^{9\,w}\, \Big\{r^6\,\Big\{6\, M\,\Big(2 + (4 - 3\, w)\, w\Big)+ r\, (-1 + 9\, w^2)\Big\}+3\,r^3\,\Big\{-36\, M^2 + 16\, M\, r - r^2 + 24\, M\, (-2\,M + r)\, w\nonumber\\
      &&+9\,(-2\, M + r)^2\,w^2\Big\}\,\xi^2
      +6\,(-2\, M + r)\,\Big\{r^2\,(1 + w)\,(-7 + 3\, w)+2\, M\, r\,\left(21 + 2\, (7 - 3\,w)\, w\right)\nonumber\\
      &&+ 4\,M^2\, \left(-16 + w\, (-10 + 3\, w)\right)\Big\}\,\xi^4\Big\}\Bigg].\label{dd7}
\end{eqnarray}

\begin{figure}
    \centering
    \includegraphics[width=0.5\linewidth]{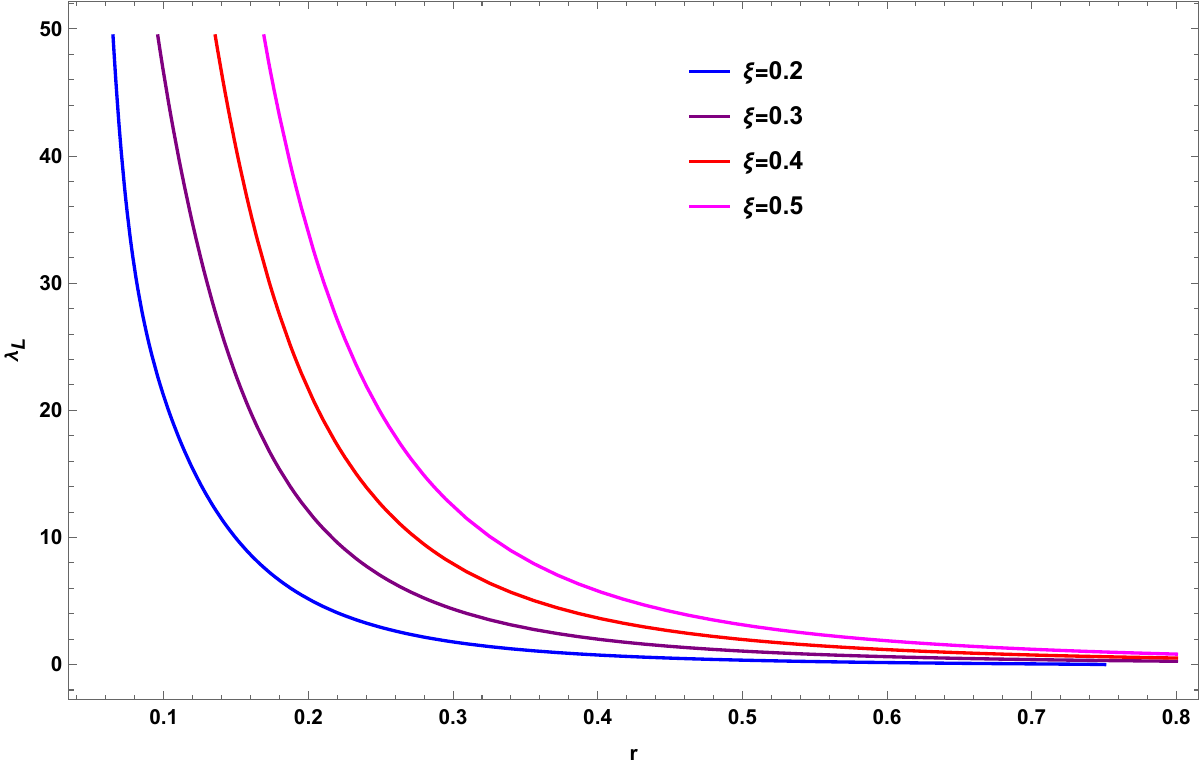}
    \caption{Illustration of the Lyapunov exponent as a function of $r$ varying quantum correction parameter $\xi$. Here $M=0.02$, $c=0.01$, $w=-2/3$.}
    \label{fig:Lyapunov}
\end{figure}

In Fig. \ref{fig:Lyapunov}, we illustrate the Lyapunov exponent as a function of the radial coordinate $r$. We observe that for specific values of the quantum correction parameter $\xi$, the normalization constant of the QF $c=0.01$, and a fixed state parameter $w = -2/3$, the Lyapunov exponent gradually decreases as $r$ increases. Furthermore, this decreasing trend shifts upward as the value of the parameter $\xi$ increases, indicating that the quantum correction parameter influences the behavior of the Lyapunov exponent function.  

In the limit where $\xi=0$, with no quantum correction, the Lyapunov exponent from Eq. (\ref{dd7}) becomes 
\begin{eqnarray}
    \lambda^2_{L}=\frac{1}{2}\,r^{-4 - 
  6\, w}\,\Big[2\,M\,(6\,M-r)\,r^{6\,w}+3\,c^2\,(1 + w)\,(1 + 3\,w)+c\,r^{3\,w}\,\left\{6\,M\,(2+4\,w-3\,w^2)-r+9\,r\,w^2\right\}\Big].\label{dd8}
\end{eqnarray}

Moreover, in the limit where $c=0$, with no QF, the Lyapunov exponent from Eq. (\ref{dd7}) becomes
\begin{eqnarray}
    \lambda_{L}=\frac{\sqrt{M\,(6\,M - r)\,r^6 - 3\, M\, r^3\, (-6\, M + r)\, (-2\, M + r)\, \xi^2 + 
 2\,(-2\, M + r)^2\, (24\, M^2 - 13\, M\, r + 2\, r^2) \xi^4}}{r^5}.\label{dd9}
\end{eqnarray}

By comparing Eqs. (\ref{dd7}), (\ref{dd8}), and(\ref{dd9}), it is evident that the Lyapunov exponent is modified by the combined effects of the quantum correction characterized by $\xi$ and the QF of parameters $(c, w)$, which together influence the gravitational field produced by the space-time geometry described in (\ref{aa1}).

\section{Periodic orbits in the EQG BH surrounded by QF}\label{sec5}

This section examines periodic orbits around the effective quantum corrected BH surrounded by quintessential field. These stable periodic orbits are crucial for kinematic analysis, determining the motion trajectories of objects and simplifying their study. In this section, we analyze and present representative periodic orbits together with their properties. Considering spherically symmetric BH spacetime and equatorial plane motion ($\theta = \pi/2$), these orbits are effectively referred to as two-dimensional.
\begin{figure}[ht!]
    \centering
    \includegraphics[width=0.45\linewidth]{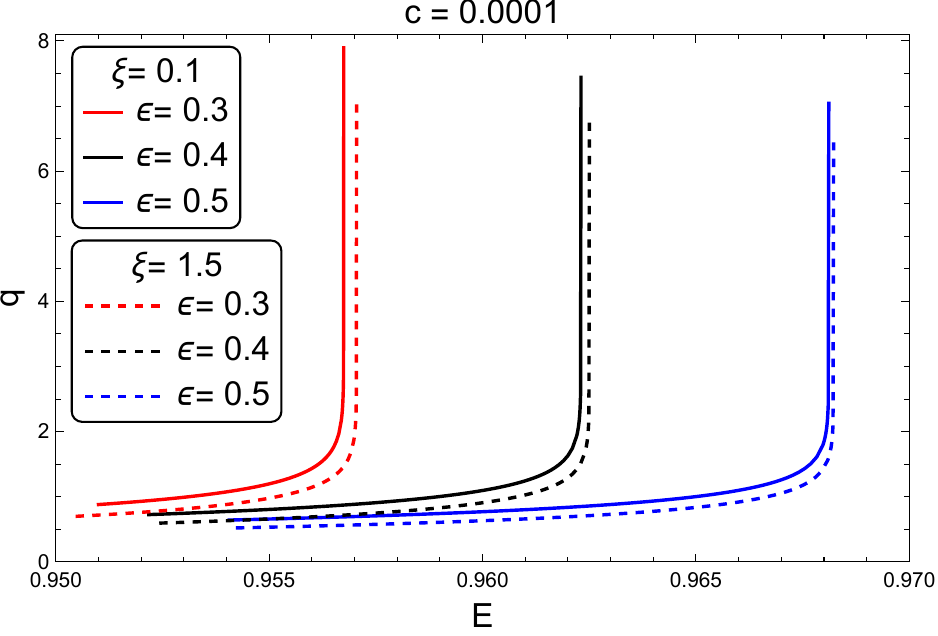}\quad\quad
    \includegraphics[width=0.45\linewidth]{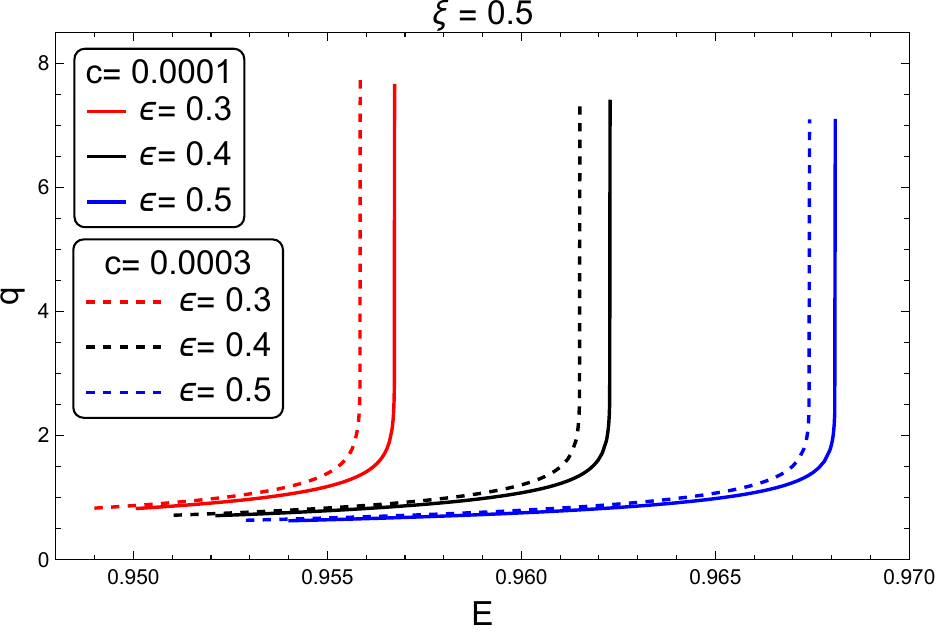}
    \caption{Illustration of the rational number $q$ as a function of energy $E$ for various combinations of quantum correction parameter $\xi$ and $\epsilon$ (left panel) and QF $c$ and $\epsilon$ (right panel). Here, we note that we have set the equation of state parameter as $w=-2/3$.}
    \label{fig:rational}
\end{figure}
    \renewcommand{\arraystretch}{1.5}
\begin{table}[]
\resizebox{1.0\textwidth}{!}{
\begin{tabular}{|c|c|c|c|c|c|c|c|c|c|c|}
\hline
$c$                     & $\xi/M$ & $L/M$   & $E_{(1,1,0)}$ & $E_{(1,2,0)}$ & $E_{(2,1,1)}$ & $E_{(2,2,1)}$ & $E_{(3,1,2)}$ & $E_{(3,2,2)}$ & $E_{(4,1,3)}$ & $E_{(4,2,3)}$ \\ \hline
\multirow{3}{*}{0.0001} & 0.1     & 3.73134 & 0.964969    & 0.968049    & 0.967678    & 0.968103    & 0.967885    & 0.968107    & 0.967948    & 0.968109    \\ \cline{2-11} 
                        & 0.3     & 3.72123 & 0.965034    & 0.968044    & 0.967683    & 0.968096    & 0.967884    & 0.968099    & 0.967945    & 0.968101    \\ \cline{2-11} 
                        & 0.5     & 3.70110 & 0.965163    & 0.968035    & 0.967694    & 0.968083    & 0.967885    & 0.968087    & 0.967942    & 0.968088    \\ \hline
\multirow{3}{*}{0.0003} & 0.1     & 3.73244 & 0.963941    & 0.967379    & 0.966957    & 0.967441    & 0.967193    & 0.967446    & 0.967264    & 0.967447    \\ \cline{2-11} 
                        & 0.3     & 3.72234 & 0.964012    & 0.967374    & 0.966963    & 0.967433    & 0.967192    & 0.967437    & 0.967262    & 0.967439    \\ \cline{2-11} 
                        & 0.5     & 3.70224 & 0.964153    & 0.967365    & 0.966977    & 0.967419    & 0.967194    & 0.967423    & 0.967259    & 0.967425    \\ \hline
\multirow{3}{*}{0.0005} & 0.1     & 3.73350 & 0.962677    & 0.966693    & 0.966156    & 0.966772    & 0.966452    & 0.966778    & 0.966543    & 0.966780    \\ \cline{2-11} 
                        & 0.3     & 3.72342 & 0.962755    & 0.966687    & 0.966165    & 0.966764    & 0.966454    & 0.966769    & 0.966542    & 0.966772    \\ \cline{2-11} 
                        & 0.5     & 3.70334 & 0.962910    & 0.966679    & 0.966185    & 0.966749    & 0.966459    & 0.966755    & 0.966543    & 0.966757    \\ \hline
\end{tabular}
}
\caption{The periodic orbits, characterized by various $(z, \omega, v)$ configurations, are defined by appropriate energies $E$ of a particle  around the EQG BH surrounded by QF for various combinations of $\xi$ and $c$ parameter. Here, we note that the scale factor and the equation of state parameter have been set $\epsilon = 0.5$ and $w=-2/3$, respectively. 
}
\label{table:periodic}
\end{table}

\begin{figure}[ht!]
    \centering
    \includegraphics[width=0.4\linewidth]{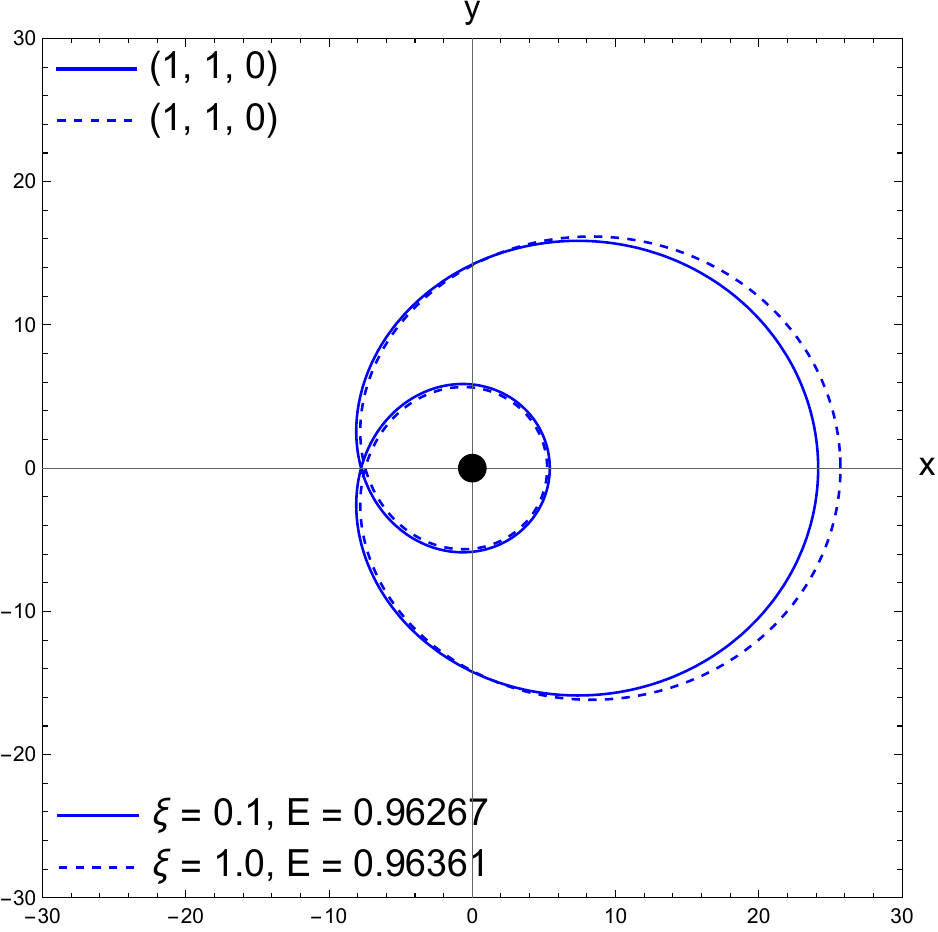}\quad\quad
    \includegraphics[width=0.4\linewidth]{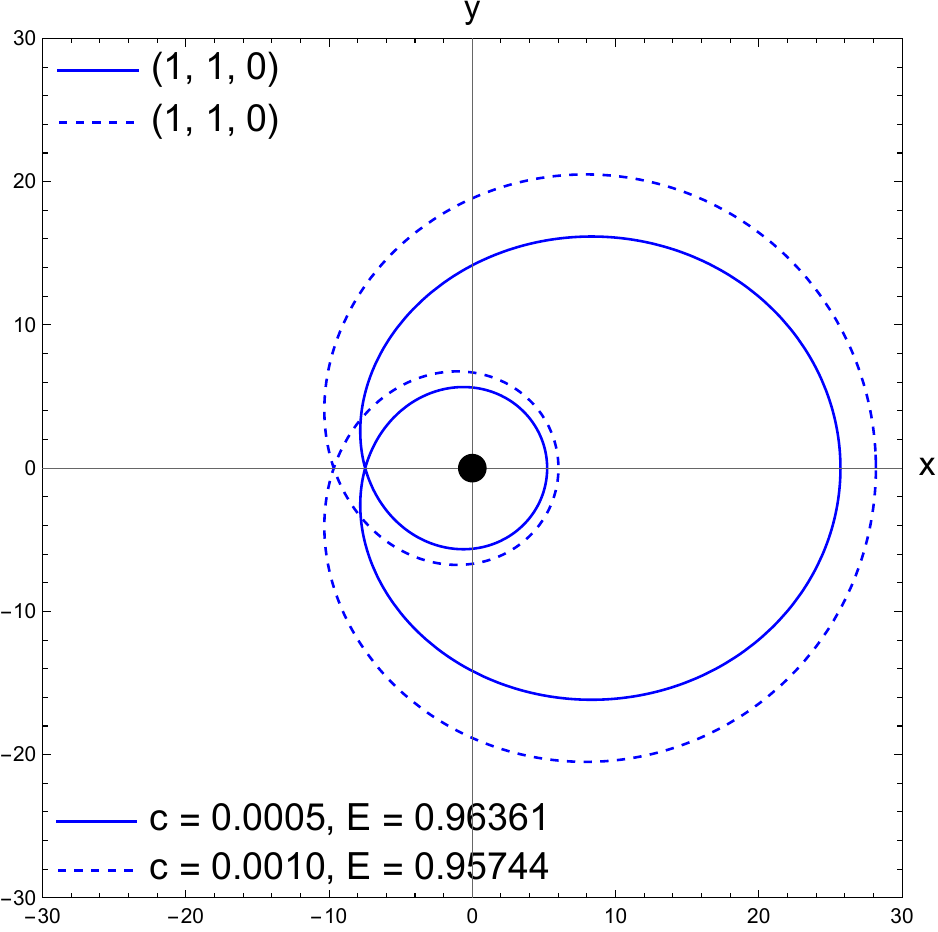}
    \caption{Illustration of periodic orbits for different values of $\xi$ and $c$ parameter with the single configuration of $(z, \omega, v)$ around the EQG BH surrounded by QF. Here, we have set $\epsilon = 0.5$ and $w=-2/3$, respectively. }
    \label{fig:comp}
\end{figure}
\begin{figure*}
\centering
\includegraphics[width=5.25cm]{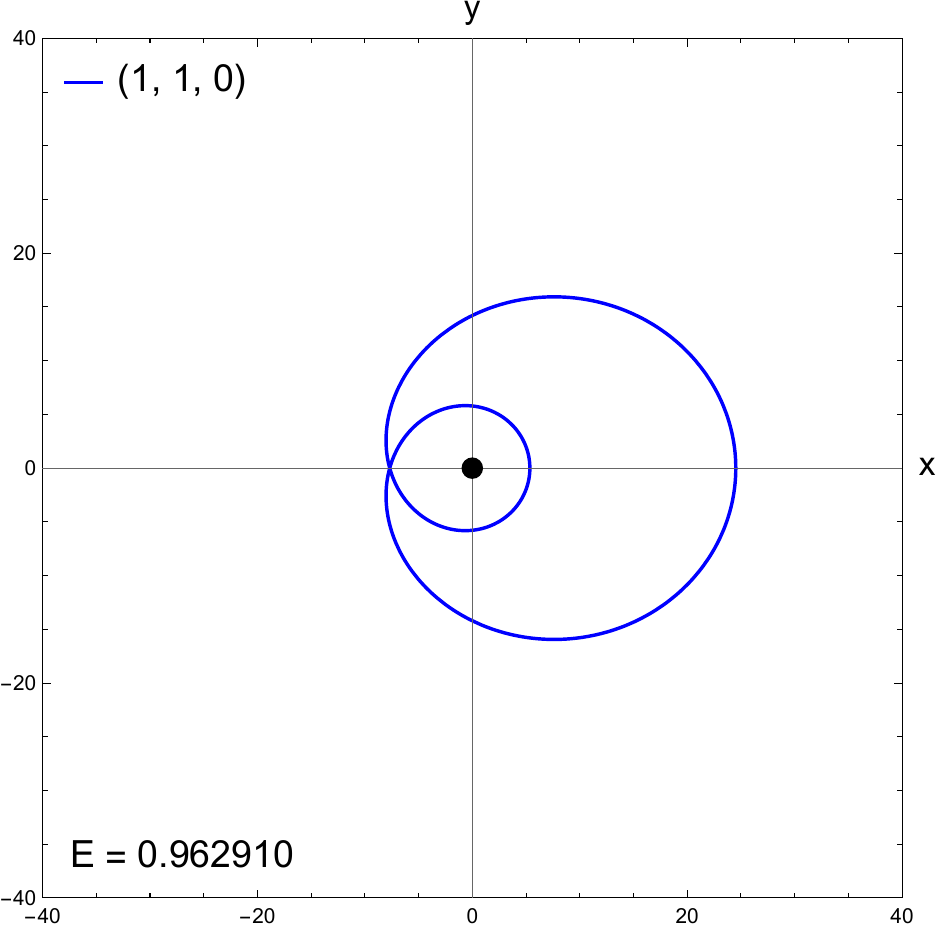} 
\includegraphics[width=5.25cm]{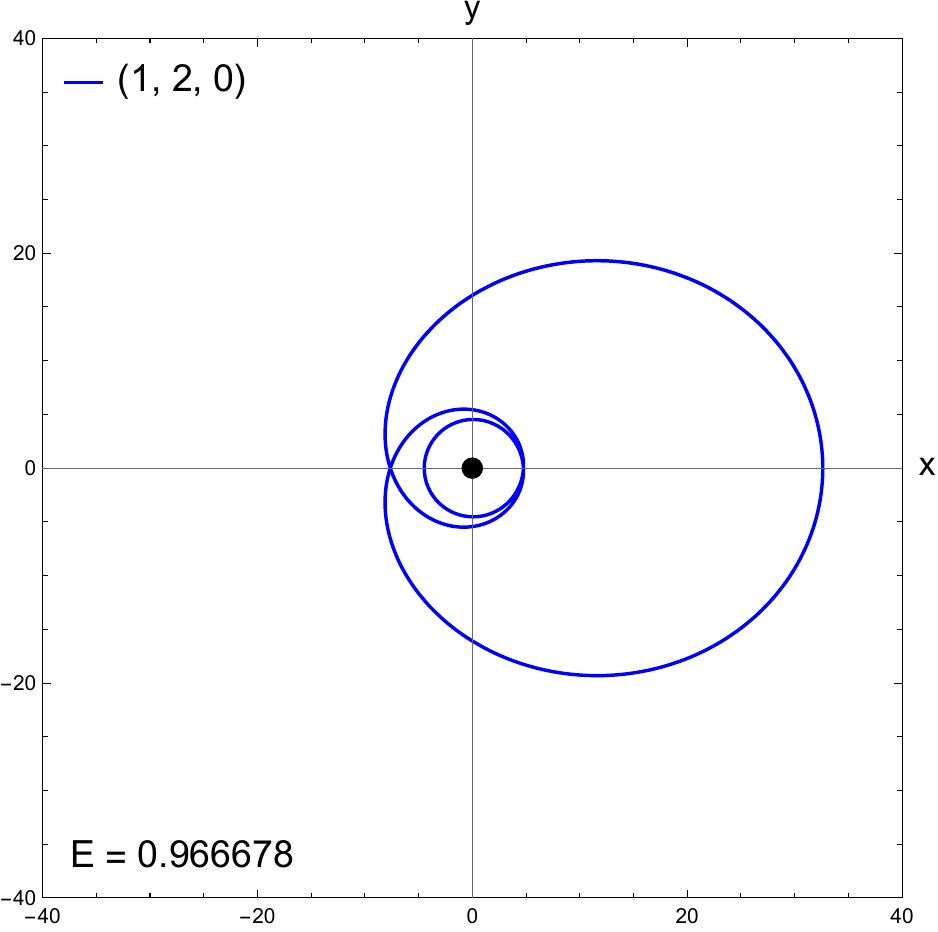}
\includegraphics[width=5.25cm]{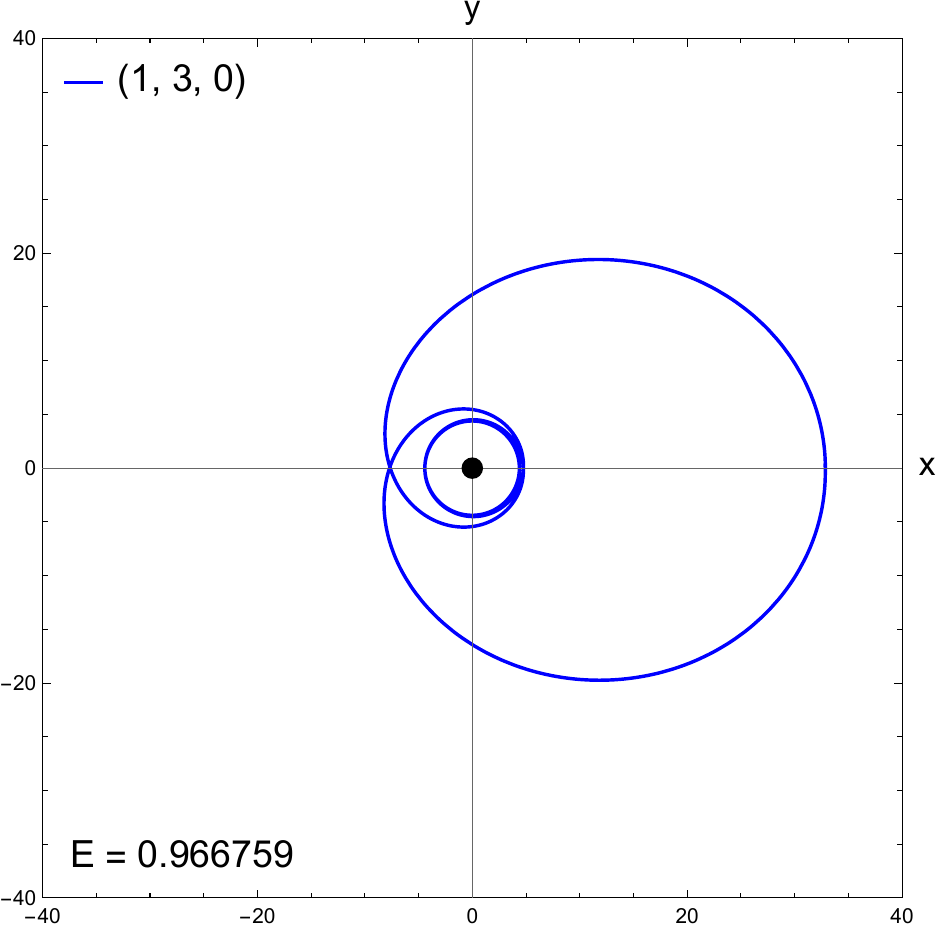} 
\includegraphics[width=5.25cm]{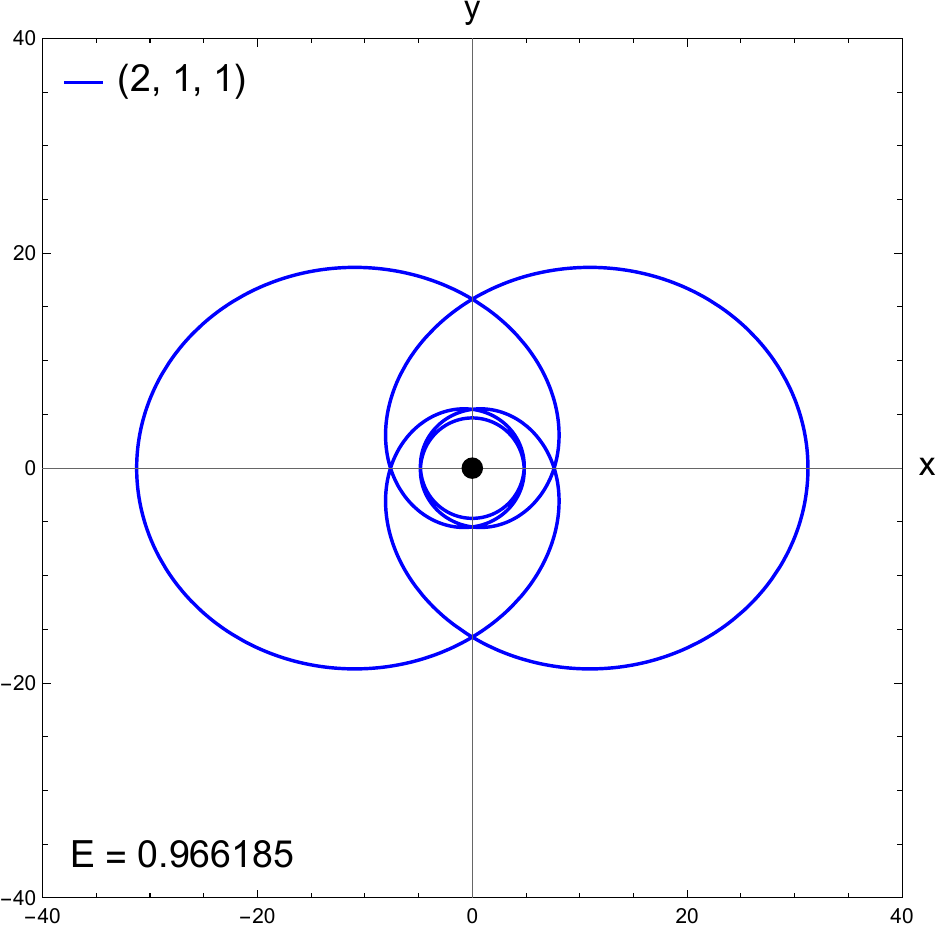} 
\includegraphics[width=5.25cm]{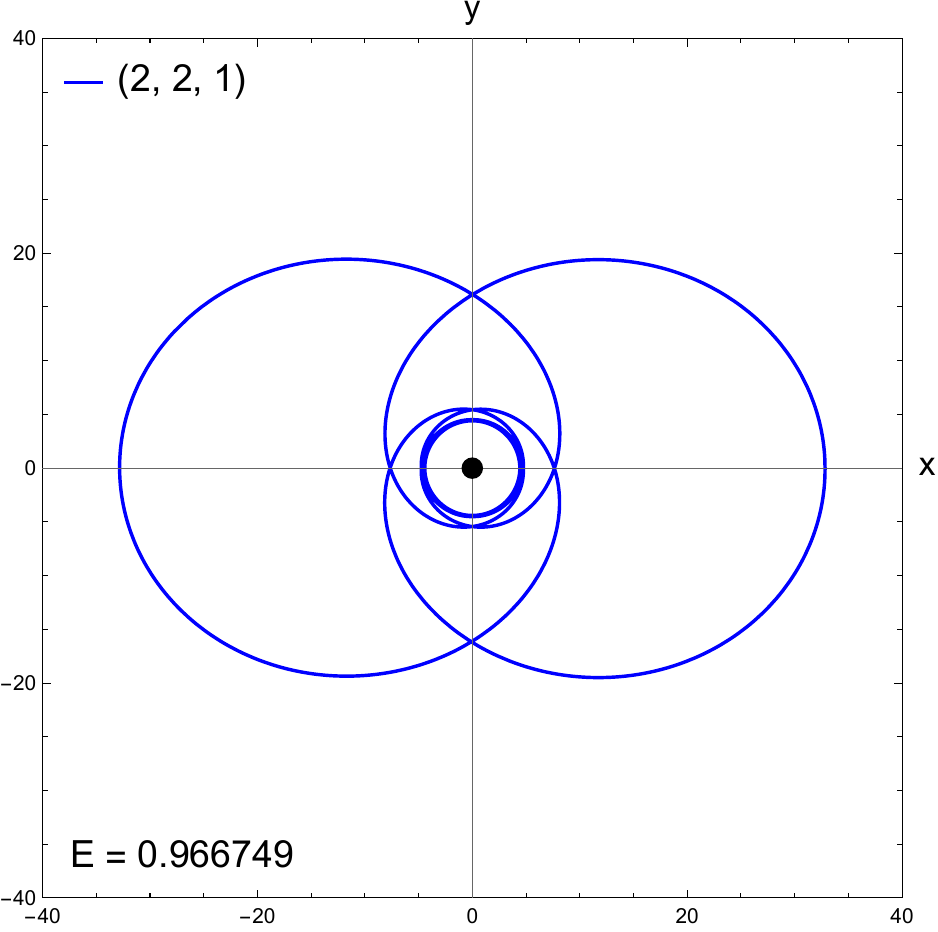} 
\includegraphics[width=5.25cm]{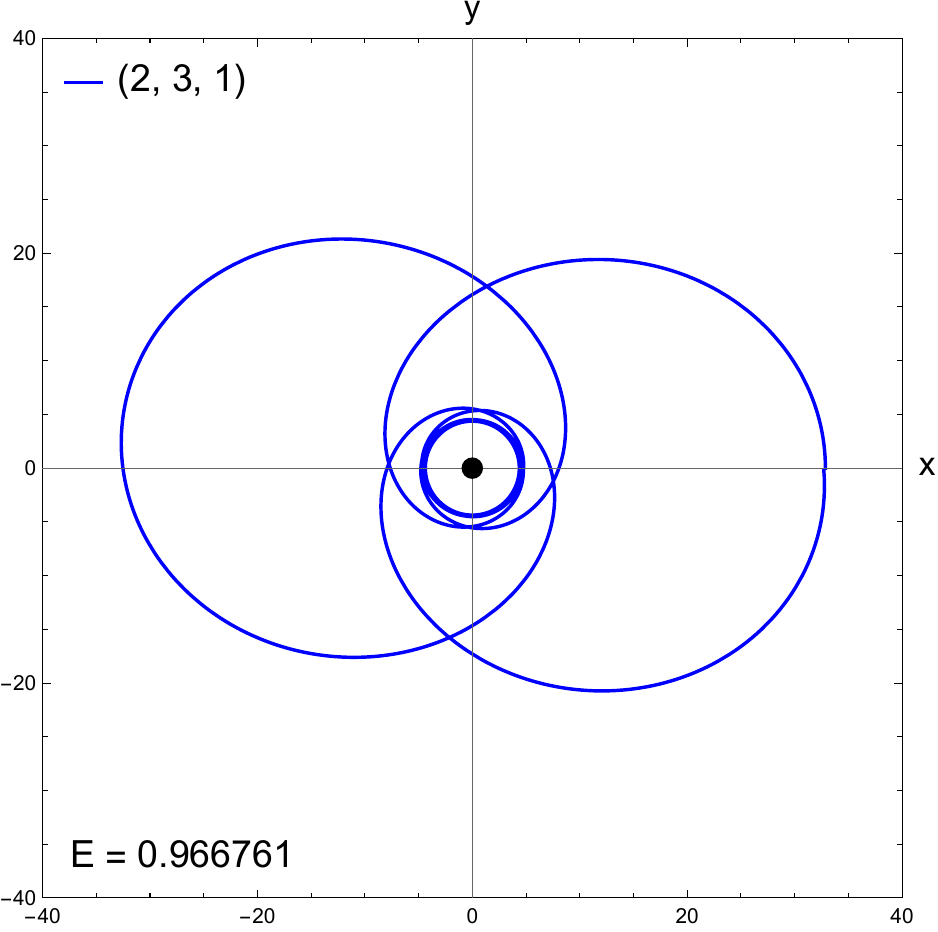} 
\includegraphics[width=5.25cm]{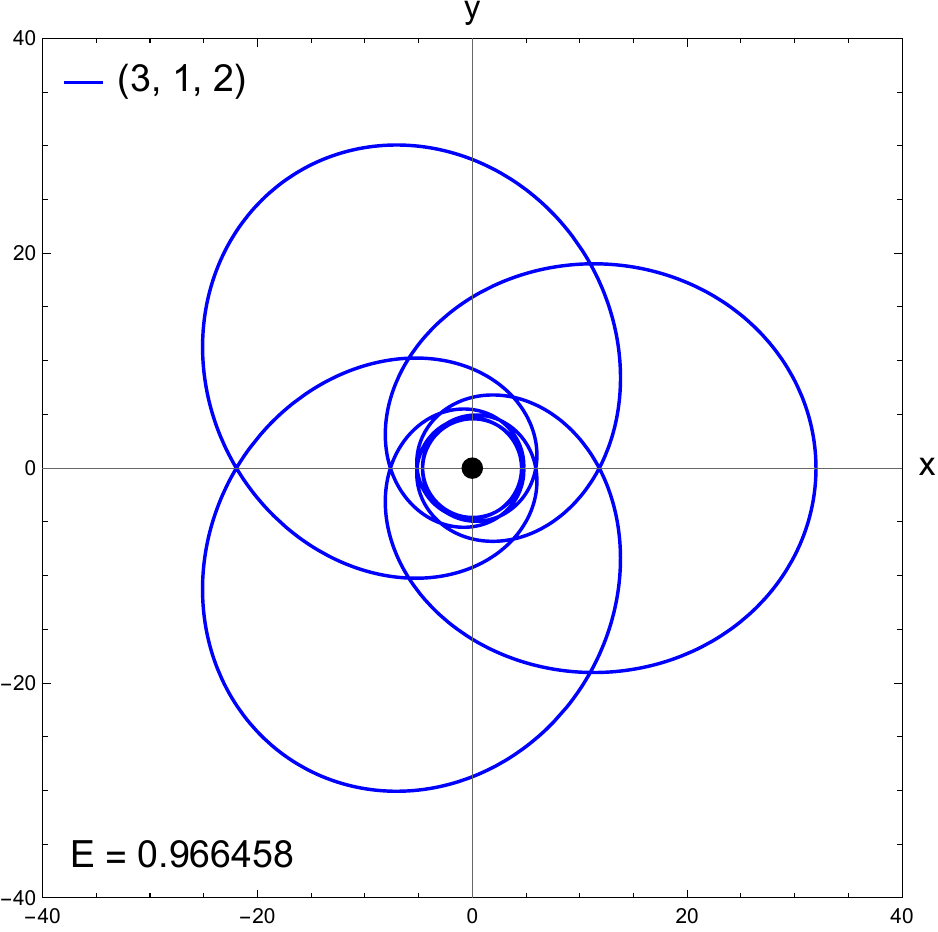} 
\includegraphics[width=5.25cm]{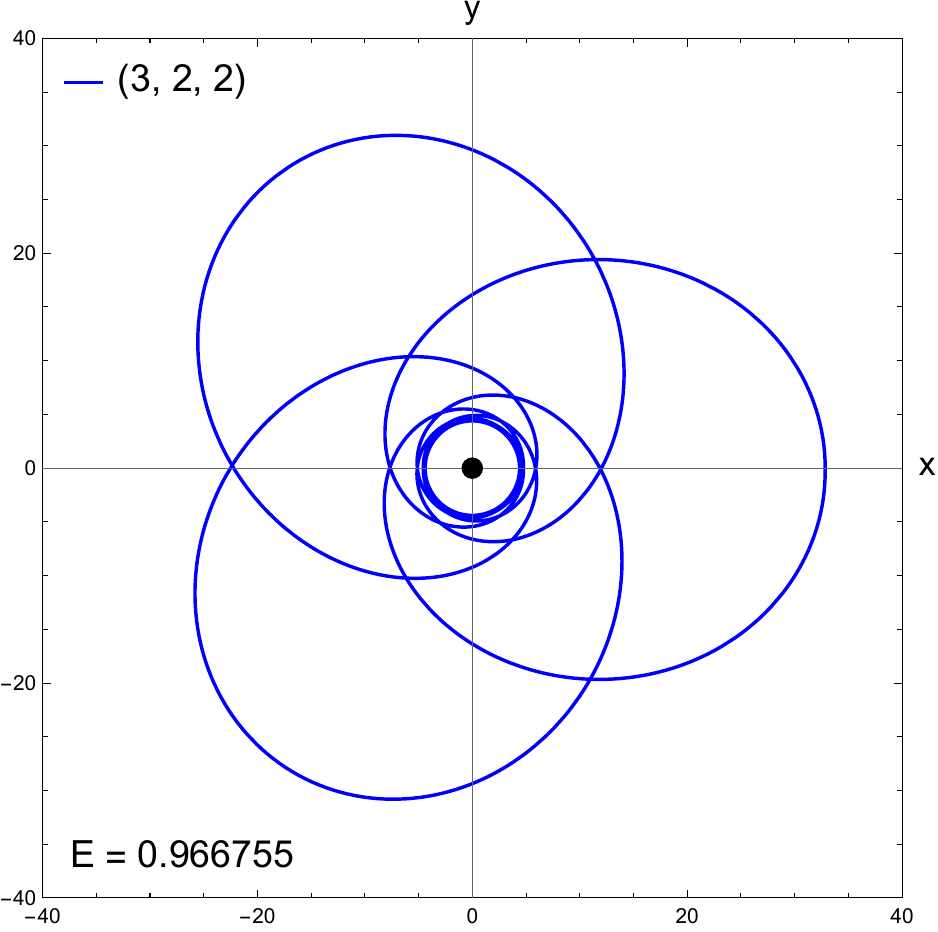} 
\includegraphics[width=5.25cm]{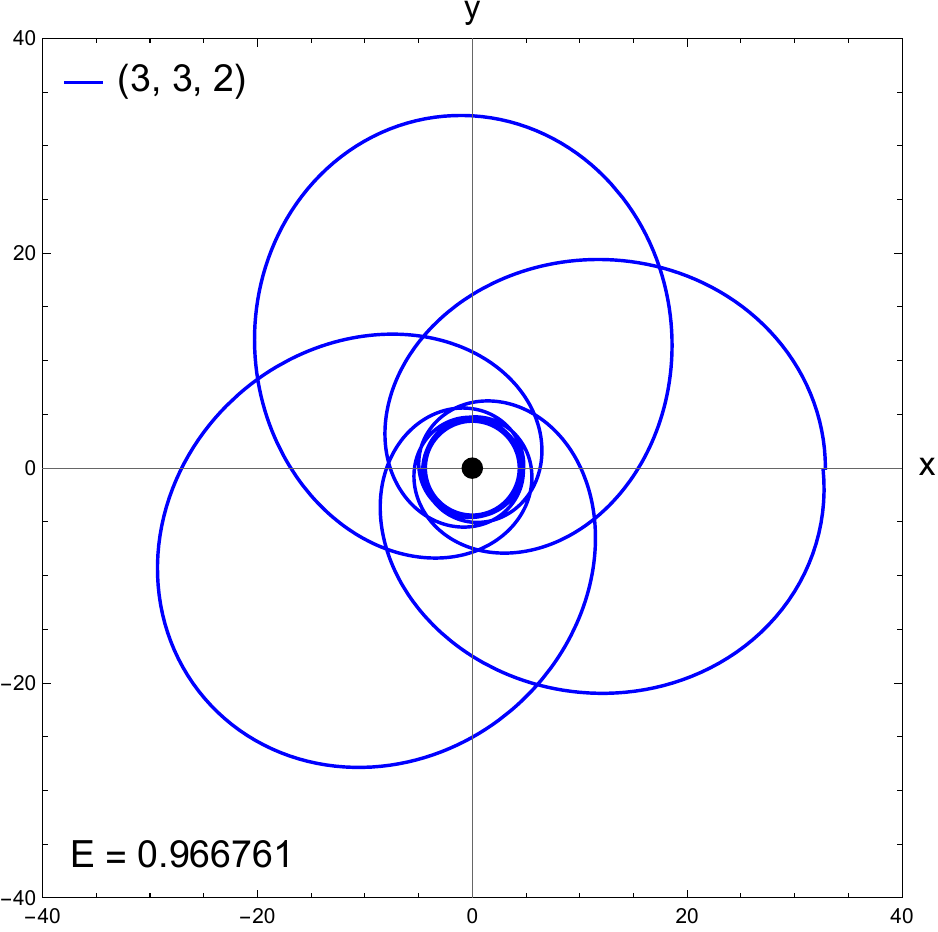} 
\includegraphics[width=5.25cm]{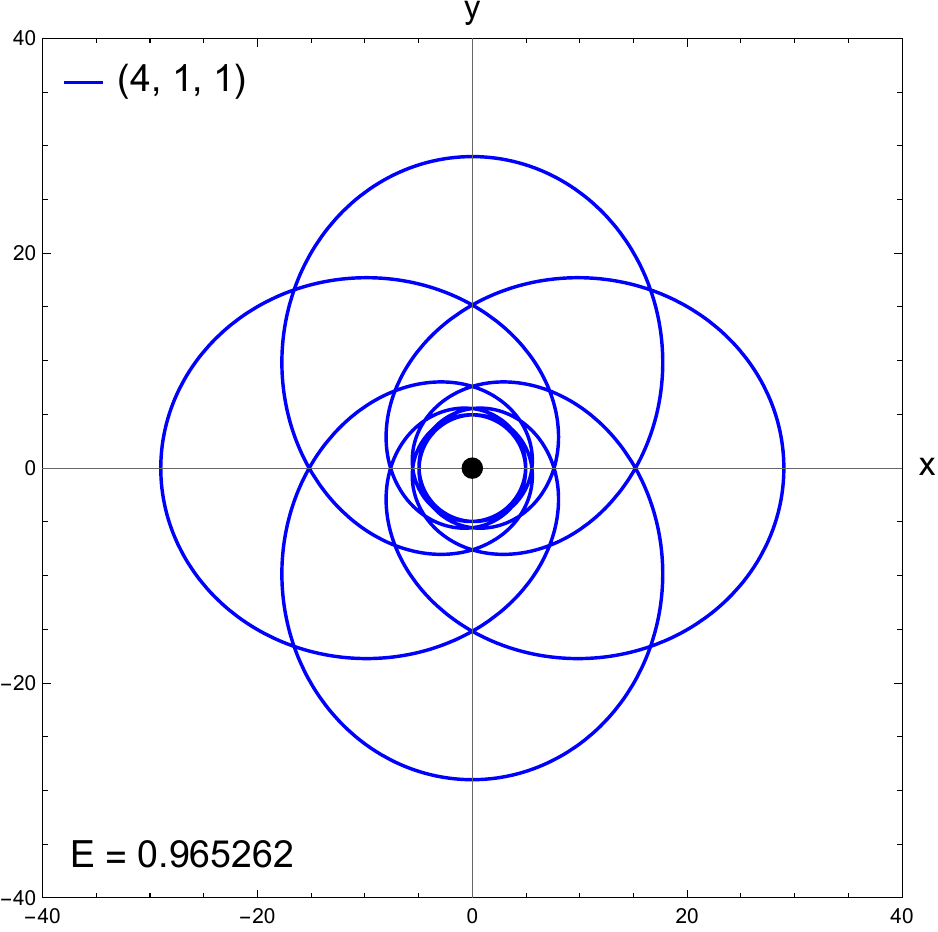} 
\includegraphics[width=5.25cm]{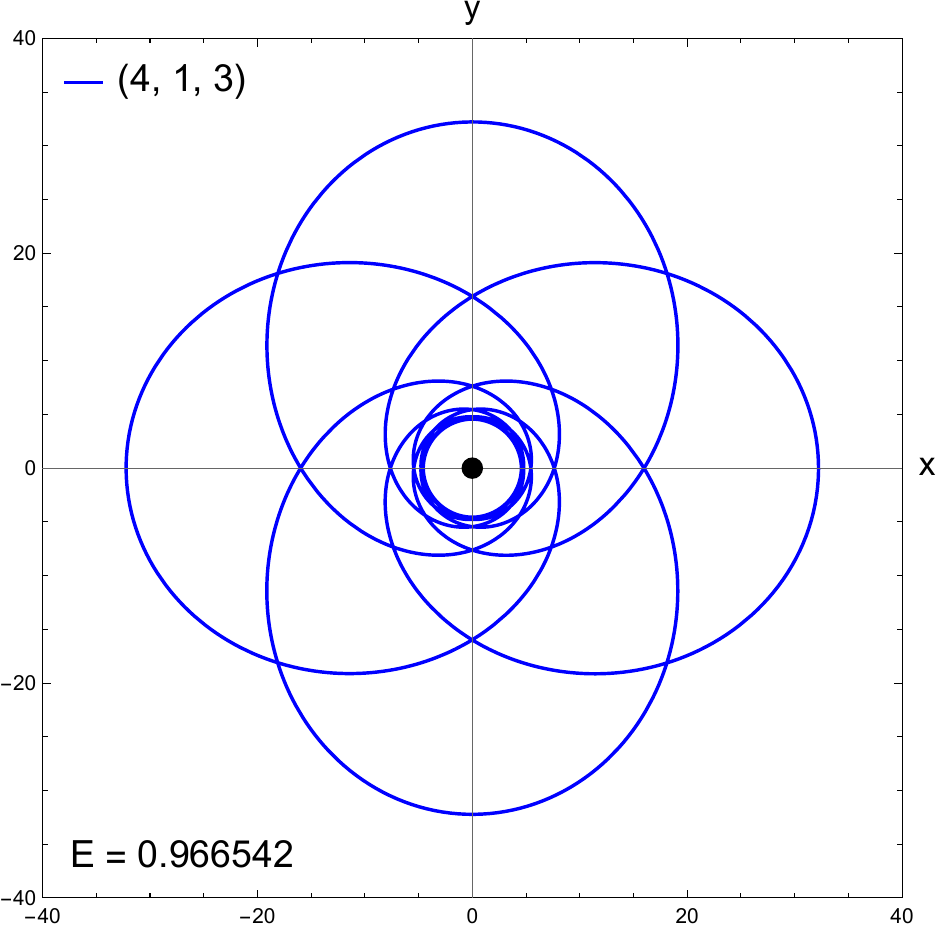} 
\includegraphics[width=5.25cm]{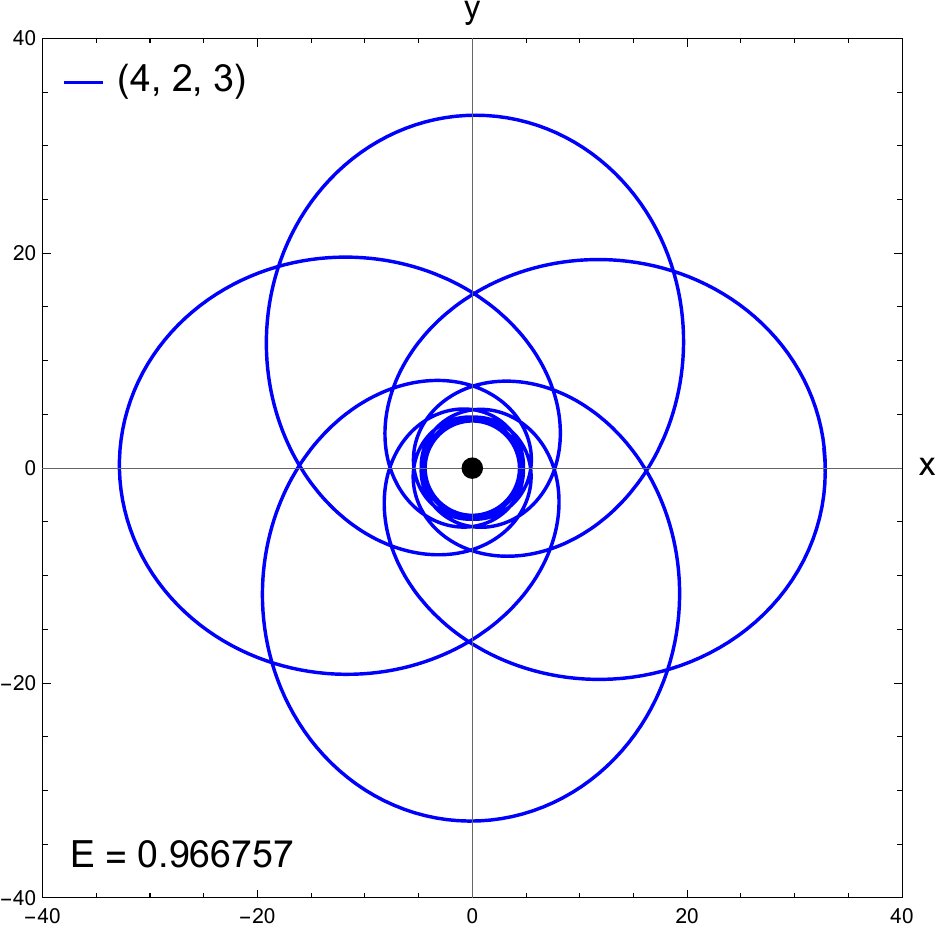} 
\captionsetup{justification=raggedright,singlelinecheck=false}
\caption{Illustration of periodic orbits for different configuration of $(z, \omega, v)$ for corresponding energies of particle around the EQG BH surrounded by QF. Here, we have set parameters as $\epsilon = 0.5$, $w=-2/3$, $\xi=0.5$, and $c=0.0005$. }
\label{fig:per_orbits}
\end{figure*}
Here, we focus on analyzing the behavior of periodic orbits using marginally bound orbits (MBO) and ISCO. It is well-established that MBO can be determined by 
    \begin{eqnarray}
        \dot{r}^2={E}^{2}-{V}_{eff}=0\mbox{\,\,\,and\,\,\,}\frac{dV_{eff}}{dr}=0,
        \label{MBO}
    \end{eqnarray}
with $E=1$. We further explore the above equations numerically, which yield values for the radius $r_{mob}$ and the angular momentum $L_{mob}$ of the marginally bound orbits. Similarly, the ISCO can also be determined by the conditions:
    \begin{equation}
        \dot{r}^2=E^2-V_{eff}=0,\,\,\,\frac{dV_{eff}}{dr}=0 \mbox{\,\,\,and\,\,\,} \frac{d^2V_{eff}}{dr^2}=0.
    \label{ISCO}
    \end{equation}

For a test particle to be on a stable orbit around a BH, the angular momentum $L$ of that particle must have particular values satisfying the allowed range given by  
\begin{equation}
        L\in[L_{isco},L_{mbo}]\, .
    \label{L-inequality}
    \end{equation}
Hence, it allows the angular momentum $L$ to be determined by the linear combination of the angular momentum on the ISCO and MBO, that is, $L_{isco}$ and $L_{mbo}$,
\begin{equation}
        L=L_{isco}+\epsilon(L_{mbo}-L_{isco})\, ,
    \label{L-linear}
    \end{equation}
with a dimensionless parameter $\epsilon$ that varies from $0$ to $1$. Analysis of the MBO and ISCO reveals that, for given $\xi$ and $c$, a stable periodic orbit around the BH requires two critical orbits: MBO and ISCO, characterized by angular momenta $L_{mbo}$ and $L_{isco}$, respectively.  A stable periodic orbit has both a perihelion and an aphelion, satisfying the condition $\dot{r}^2(r)=0$ with three roots. Notably, the smallest of these three roots is irrelevant to our study. Thus, we retain only the other two roots, representing the perihelion and aphelion, which must satisfy the following inequality for a bound orbit to exist
    \begin{equation}
        \dot{r}^2(r_{1})\,\dot{r}^2(r_{2})<0,
        \label{r2-inequality}
    \end{equation}
or 
    \begin{equation}
        V_{eff}(r_{2})<E^{2}<V_{eff}(r_{1})\, ,\,\, r_{1}<r_{2}\, ,
        \label{Value range of E}
    \end{equation}
where $r_{1}$ and $r_{2}$ are the radial coordinates and represent two extreme points of $\dot{r}^2(r)$ defining the existence of stable orbits. 

We then turn to defining the periodic orbits, which occur between the ISCO and MBO and can be represented by a rational number $q$ defined by 
    \begin{equation}
        q=\omega+\frac{v}{z}=\frac{\Delta\phi}{2\pi}-1\, .
        \label{q}
    \end{equation}
Additionally, we note that the rational number $q$ is represented by three integers $(z, \omega, v)$, that is, zoom $z$, whirl $\omega$, and vertex $v$, satisfying $v<z$. Here, $\Delta\phi$ refers to as a numerical integral, which is defined by 
    \begin{equation}
        \Delta\phi=\oint d\phi\, .   
        \label{Deltaphi}
    \end{equation}
We further explore it numerically, focusing on analyzing integration $\Delta\phi$, which is determined by
    \begin{equation}
        \begin{aligned}
            \Delta\phi&=2\int_{\phi_{1}}^{\phi_{2}}d\phi=2\int_{r_{per}}^{r_{aph}}\frac{L}{r^{2}\sqrt{E^2-V_\text{eff}(r)}}\,dr\, ,
        \end{aligned}
        \label{numerical integral}
    \end{equation}
and represents the equatorial angle accumulated over a radial period, from one perihelion to the aphelion, and then to the next perihelion. The limits of the integral, $r_{per}$ and $r_{aph}$, correspond to the two turning points of a bound orbit between the ISCO and the MBO. These points satisfy the condition
\begin{equation}
        \dot{r}^{2}(r)=0\, .
        \label{rsqurae=0}
    \end{equation}
As conditions for the existence of bound orbits have been outlined earlier, the particle energy $E$ must satisfy Eq.~(\ref{Value range of E}) with the two largest roots corresponding to $r_{per}$ and $r_{aph}$.

We now analyze the relation between the energy of a particle and the rational number $q$ representing the periodic orbits around the BH. In Fig.~\ref{fig:rational}, we demonstrate the energy profile of the rational number $q$ for different values of the parameters $\epsilon$, $\xi$ and $c$. It can be seen from Fig.~\ref{fig:rational} that the curves move to the right towards larger values of the energy with increasing parameter $\epsilon$. The periodic orbits around the BH can only exist at the largest values of the particle energy when increasing the parameter $\epsilon$. Similarly, one can observe that the curve slightly shifts to the right as a consequence of an increase in the value of the quantum correction parameter $\xi$ in correspondence with the behavior of the $\epsilon$ and $\xi$, the quintessential parameter $c$ inversely affects the behavior of the functional relationship between $q$ and $E$, i.e., the curve slightly shifts to the left towards smaller values of the energy as the parameter $c$ increases, resulting in periodic orbits occurring at small energy of the particle around the BH, as seen in the right panel of Fig.~\ref{fig:rational}.  
Additionally, it is worth noting that the growth rate of $q$ decreases as the energy $E$ gradually reaches its maximum $E_{max}$, tending to its smaller number. For the existence of stable orbits in correspondence with the appropriate values of parameters $\xi$ and $c$, the MBO and the ISCO are key requirements, satisfying the rational number $q$ that is allowed to have values in the following range
    \begin{equation}
    q(E_{min})\leqslant q\leqslant q(E_{max})\, ,
        \label{q value range}
    \end{equation}  
together with the following required conditions 
    \begin{equation}
        \dot{r}^2(r_{per},E_{min})=\dot{r}^2(r_{aph},E_{max})=0\, .
        \label{E value range}
    \end{equation}
    However, Fig.~\ref{fig:rational} suggests that $q(E_{max})$ can become arbitrarily large under certain conditions, such as in an extremely strong gravitational field regime.

 To enhance our understanding of the behavior of the periodic orbits, we consider several examples using the rational number $q$ and various combinations of integers $(z, \omega, v)$. In Fig.~\ref{fig:comp} and \ref{fig:per_orbits}, we show simple periodic orbits with their corresponding rational numbers and integers. It must be emphasized that the rational number $q$ describes the orbit's geometric characteristics, while specifically, $z$ represents the number of loops in the orbit, $\omega$ the number of rotations around the BH, and $v$ the order in which the loops are traversed with complex trajectories. Fig.~\ref{fig:comp} illustrates how the impacts of parameters $\xi$ and $c$ change the behavior of the periodic orbits, particularly for the chosen combinations of integers, e.g., $(1, 1, 0)$.  It can be seen from Fig.~\ref{fig:comp} that the orbit exhibits periodic behavior. It is initially almost stable around the BH, but it begins to become larger as the quantum correction parameter $\xi$ gradually increases at larger distances, as seen in the left panel of Fig.~\ref{fig:comp}. Unlike the quantum correction effect, the orbit exhibits similar behavior, but becomes larger at both small and larger distances under the impact of the quintessential field parameter $c$, as shown in the right panel. In Fig.~\ref{fig:per_orbits}, periodic orbits are shown and collectively referred to as the general case for various combinations of integers $(z, \omega, v)$ for fixed corresponding values of $\epsilon = 0.5$, $\xi=0.5$, and $c=0.0005$. From Fig.~\ref{fig:per_orbits}, the number of loops and rotations in periodic orbits increases as $z$ and $\omega$ increase with complex trajectories. Furthermore, a detailed analysis of the periodic orbits around the BH is reflected in Table~\ref{table:periodic}. It can be seen from Table~\ref{table:periodic} that the energy of the particle requires slightly larger values in the periodic orbits as a consequence of the impact of the quantum correction parameter $\xi$. This occurs because $\xi$ enhances the background gravity that can affect the behavior of the periodic orbits around the EQG BH surrounded by QF. 

\section{Scalar Perturbations} \label{sec6}

In this section, we explore the dynamics of a massless scalar field in the background of an effective quantum-corrected BH solution surrounded by a QF. We begin by deriving the massless Klein-Gordon equation, which governs the evolution of the scalar field within the specified space-time geometry.

Scalar perturbations in BH space-times are essential for understanding BH stability. These perturbations have been extensively studied in various BH solutions within general relativity, offering valuable insights into both the stability of BHs and the propagation of fields in curved space-time. For example, scalar field perturbations have been analyzed in Schwarzschild, Kerr, and Reissner-Nordström BHs, as well as in other BH solutions within general relativity and modified gravity theories (see, {\it e.g.}, Refs. \cite{NPB, CJPHY, EPJC}, and related works).

The massless scalar field wave equation is described by the Klein-Gordon equation as follows Refs. \cite{NPB,CJPHY,EPJC}:
\begin{equation}
\frac{1}{\sqrt{-g}}\,\partial_{\mu}\left[\left(\sqrt{-g}\,g^{\mu\nu}\,\partial_{\nu}\right)\,\Psi\right]=0,\label{ff1}    
\end{equation}
where $\Psi$ is the wave function of the scalar field, $g_{\mu\nu}$ is the covariant metric tensor, $g=\det(g_{\mu\nu})$ is the determinant of the metric tensor, $g^{\mu\nu}$ is the contrvariant form of the metric tensor, and $\partial_{\mu}$ is the partial derivative with respect to the coordinate systems.

Before, writing explicitly, performing the following coordinate change (called tortoise coordinate) 
\begin{eqnarray}
    dr_*=\frac{dr}{\mathcal{F}(r)}\label{ff2}
\end{eqnarray}
into the line-element Eq. (\ref{aa1}) results
\begin{equation}
    ds^2=\mathcal{F}(r_*)\,\left(-dt^2+dr^2_{*}\right)+\mathcal{H}^2(r_*)\,\left(d\theta^2+\sin^2 \theta\,d\phi^2\right),\label{ff3}
\end{equation}
where $\mathcal{F}(r_*)$ and $\mathcal{H}(r_*)$ are functions of $r_*$. 

Let us consider the following scalar field wave function ansatz form
\begin{equation}
    \Psi(t, r_{*},\theta, \phi)=\exp(i\,\omega\,t)\,Y^{m}_{\ell} (\theta,\phi)\,\frac{\psi(r_*)}{r_{*}},\label{ff4}
\end{equation}
where $\omega$ is (possibly complex) the temporal frequency, $\psi (r_*)$ is a propagating scalar field in the candidate space-time, and $Y^{m}_{\ell} (\theta,\phi)$ is the spherical harmonics.

With these, we can write the wave equation (\ref{ff1}) in the following form:
\begin{equation}
    \frac{\partial^2 \psi(r_*)}{\partial r^2_{*}}+\left(\omega^2-\mathcal{V}\right)\,\psi(r_*)=0,\label{ff5}
\end{equation}
where the effective potential is given by
\begin{eqnarray}
\mathcal{V}(r)&=&\left(\frac{\ell\,(\ell+1)}{r^2}+\frac{\mathcal{F}'(r)}{r}\right)\,\mathcal{F}(r)\nonumber\\
&=&\left[\frac{\ell\,(\ell+1)}{r^2}+\frac{2}{r^2}\,\left\{\frac{M}{r}+\frac{c\,(3\,w+1)/2}{r^{3\,w+1}}+\frac{\xi^2}{r^2}\,\left(1-\frac{2\,M}{r}-\frac{c}{r^{3\,w+1}} \right)\,\left(\frac{4\,M}{r}-1+\frac{c\,(3\,w+2)}{r^{3\,w+1}} \right) \right\}\right]\times\nonumber\\
&&\left[\left(1-\frac{2\,M}{r}-\frac{c}{r^{3\,w+1}}\right)+\frac{\xi^2}{r^2}\left( 1-\frac{2\,M}{r}-\frac{c}{r^{3\,w+1}}\right)^2\right].\label{ff6}
\end{eqnarray}

In the limit where $\xi=0$, that is, with no quantum correction in BH solution, the scalar perturbative potential from Eq. (\ref{ff6}) reduces as
\begin{eqnarray}
\mathcal{V}(r)&=&\left[\frac{\ell\,(\ell+1)}{r^2}+\frac{2\,M}{r^3}+\frac{c\,(3\,w+1)}{r^{3\,(w+1)}}\right]\,\left(1-\frac{2\,M}{r}-\frac{c}{r^{3\,w+1}}\right).\label{ff7}
\end{eqnarray}
Moreover, in the limit where $c=0$, with no QF, the scalar perturbative potential from Eq. (\ref{ff6}) reduces as
\begin{eqnarray}
\mathcal{V}(r)=\left[\frac{\ell\,(\ell+1)}{r^2}+\frac{2\,M}{r^3}-\frac{2\,\xi^2}{r^4}\,\left(1-\frac{2\,M}{r}\right)\,\left(1-\frac{4\,M}{r}\right)\right]\,\left\{1+\frac{\xi^2}{r^2}\left(1-\frac{2\,M}{r}\right)\right\}\,\left(1-\frac{2\,M}{r}\right).\label{ff8}
\end{eqnarray}
Thereby, comparing Eqs. (\ref{ff6}), (\ref{ff7}) and (\ref{ff8}), we observed that the scalar perturbative potential is modified by the quantum correction characterized by $\xi$ and the QF characterized by the parameters $(c, w)$, which together influence the gravitational field generated  by the space-time geometry described in (\ref{aa1}).

Now, we discuss the perturbative potential for a particular state parameter, $w=-2/3$. Therefore, from Eq. (\ref{ff6}), we find
\begin{eqnarray}
\mathcal{V}&=&\left[\frac{\ell\,(\ell+1)}{r^2}+\frac{2}{r^2}\,\left\{\frac{M}{r}-\frac{c\,r}{2}+\frac{\xi^2}{r^2}\,\left(1-\frac{2\,M}{r}-c\,r\right)\,\left(\frac{4\,M}{r}-1\right) \right\}\right]\times\nonumber\\
&&\left[\left(1-\frac{2\,M}{r}-c\,r\right)+\frac{\xi^2}{r^2}\left( 1-\frac{2\,M}{r}-c\,r\right)^2\right].\label{ff9}
\end{eqnarray}

Defining dimensionless variables $x=r/M$, $y=\xi/M$ and $\mathrm{a}=c\,M$, we define the following quantity
\begin{eqnarray}
M^2\,\mathcal{V}&=&\left[\frac{\ell\,(\ell+1)}{x^2}+\frac{2}{x^2}\,\left\{\frac{1}{x}-\frac{\mathrm{a}\,x}{2}+\frac{y^2}{x^2}\,\left(1-\frac{2}{x}-\mathrm{a}\,x\right)\,\left(\frac{4}{x}-1\right) \right\}\right]\times\nonumber\\
&&\left[\left(1-\frac{2}{x}-\mathrm{a}\,x\right)+\frac{y^2}{x^2}\left( 1-\frac{2}{x}-\mathrm{a}\,x\right)^2\right].\label{ff10}
\end{eqnarray}

For dominant mode, $\ell=0$, which correspond to $s$-wave scalar field, the perturbative potential (\ref{ff10}) reduces as
\begin{eqnarray}
M^2\,\mathcal{V}\Big{|}_{\ell=0}=\left[\frac{2}{x^3}-\frac{\mathrm{a}}{x}+\frac{2\,y^2}{x^4}\,\left(1-\frac{2}{x}-\mathrm{a}\,x\right)\,\left(\frac{4}{x}-1\right) \right]
\left[\left(1-\frac{2}{x}-\mathrm{a}\,x\right)+\frac{y^2}{x^2}\left( 1-\frac{2}{x}-\mathrm{a}\,x\right)^2\right].\label{ff11}
\end{eqnarray}

\begin{figure}[ht!]
    \centering
    \includegraphics[width=0.6\linewidth]{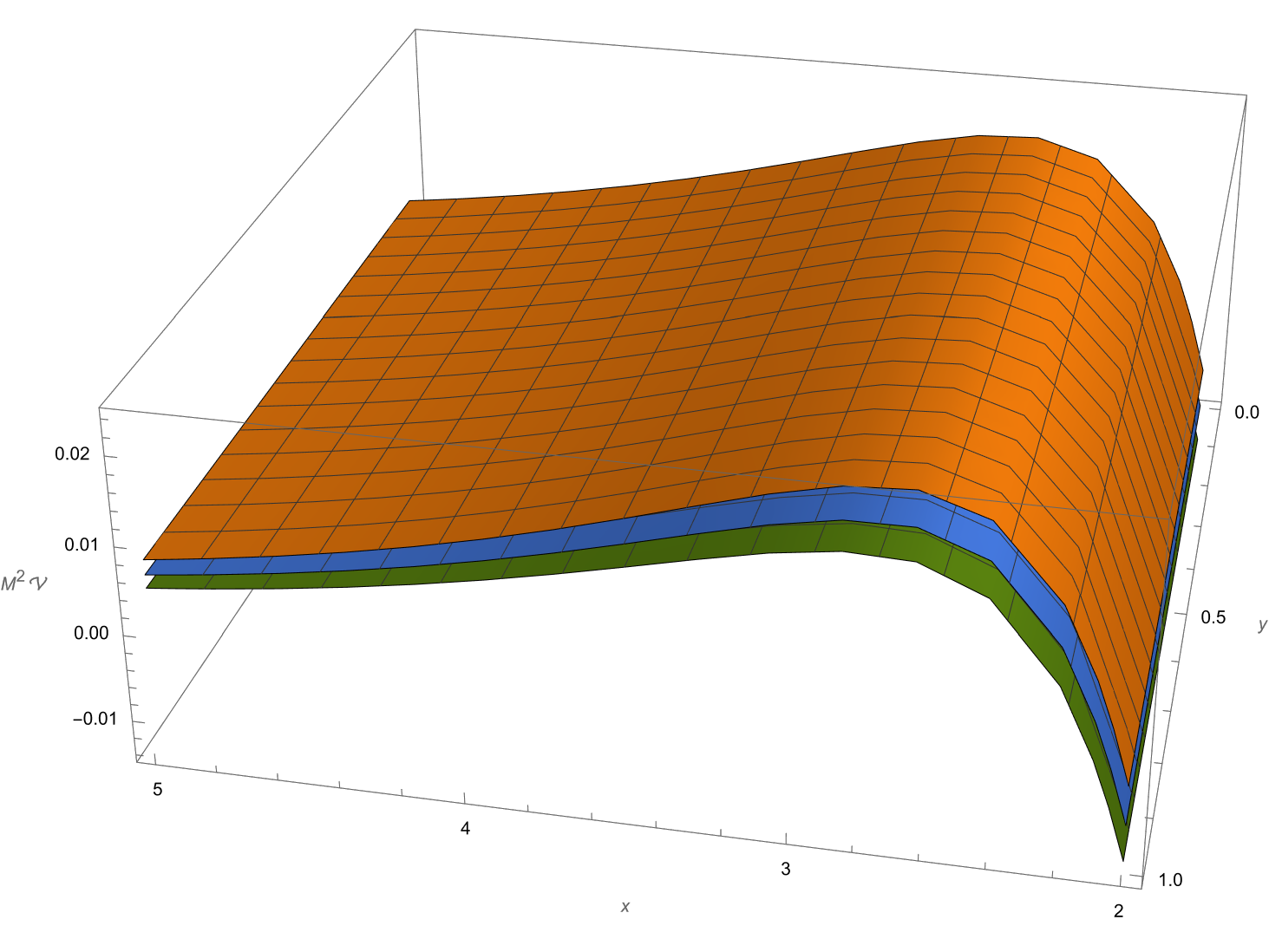}
    \caption{Three-dimensional plot of $M^2\,\mathcal{V}$: the qualitative features of zero spin scalar potential for the
 dominant multipole number $\ell=0$.}
    \label{fig:contour}
\end{figure}

In Fig. \ref{fig:contour}, we present a three-dimensional plot of the dimensionless quantities $M,\mathcal{V}$, $x$, and $y$, while varying the dimensional parameter $\mathrm{a}$ for the dominant mode $\ell=0$. The color scheme is as follows: orange represents $\mathrm{a} = 0.01$, blue corresponds to $\mathrm{a} = 0.02$, and green corresponds to $\mathrm{a} = 0.03$.

\section{Conclusions and Remarks} \label{sec7}

In this study, we investigated BHs surrounded by QF within the EQG framework. Our work focused on understanding how the interplay between quantum corrections (parameter $\xi$) and quintessence (parameters $c$ and $w$) affects observable properties and particle dynamics.

We analyzed the horizon structure of these modified BHs in Section \ref{sec2}, finding analytical solutions for three distinct horizons: the Cauchy radius $r_-$, event horizon $r_+$, and cosmological horizon $r_c$. Our analysis demonstrated that increasing the quintessence parameter $c$ enlarges the event horizon but reduces the cosmological horizon, while the quantum parameter $\xi$ primarily affects the Cauchy horizon, as visualized in Fig. \ref{figa9}. A significant contribution of our work was the derivation of analytical expressions for the photon sphere and shadow radius for various quintessence state parameters, presented in Section \ref{sec3}. These formulas, given by Eqs. (\ref{bb16c}-\ref{bb16cc}), revealed that increasing $c$ enlarges the shadow radius, while increasing $\xi$ reduces it. We also established observational constraints on $\xi$ based on Event Horizon Telescope data from M87*, finding that the allowable range for $\xi$ increases with the quintessence parameter $c$, as shown in Fig. \ref{figa8}.

In Section \ref{sec4}, we examined geodesic motions of both null and timelike particles. For photons, we derived the effective potential and force expressions, showing that both quantities increase with higher values of $\xi$. For massive particles, we investigated circular orbits, finding that increasing $\xi$ decreases angular momentum but increases orbital energy. Our analysis of the ISCO showed that both the ISCO radius and specific energy increase with $\xi$, while specific angular momentum decreases, as illustrated in Fig. \ref{fig:isco}. The stability of circular orbits was assessed through the Lyapunov exponent, which we found decreases with radial distance but shifts upward with higher values of $\xi$. This indicates that quantum corrections influence orbital stability properties, with potential implications for accretion processes around astrophysical BHs.

Our investigation of periodic orbits in Section \ref{sec5} revealed complex zoom-whirl dynamics characterized by the rational number $q$. We found that the energy profiles shift rightward with increasing $\xi$ but leftward with increasing $c$, indicating that quintessence allows periodic orbits at lower energies. The diverse orbital configurations displayed in Fig. \ref{fig:per_orbits} demonstrate how increasing zoom and whirl numbers create increasingly complex trajectories. Finally, we examined scalar perturbations in Section \ref{sec6}, deriving an effective potential that incorporates both quantum corrections and quintessence effects. The three-dimensional visualization in Fig. \ref{fig:contour} illustrates how these parameters jointly shape the potential profile, providing insights into stability and response characteristics. Throughout our analysis, we consistently found that quantum corrections and quintessence often produce opposing effects on observable quantities and particle dynamics. This interplay creates distinctive signatures that could potentially be detected by future high-precision observations, offering pathways to test dark energy models in strong-field regimes.

Several promising directions emerge for future research. Extending our analysis to include rotation would provide more realistic models for astrophysical BHs \cite{isz37,isz38}, potentially revealing additional features in shadow morphology and particle dynamics. Investigating the implications for gravitational wave signals, particularly from extreme mass-ratio inspirals \cite{isz39,isz40}, could connect our theoretical predictions with future observations. Detailed studies of accretion processes in these spacetimes \cite{isz41} would enable comparisons with electromagnetic data. Subsequently, examining thermodynamic characteristics alongside Hawking radiation \cite{isz41x1,isz41x2,isz41x3,isz42,isz43,isz44,isz44x1} could potentially highlight quantum phenomena more distinctly. These represent the forthcoming phases of our research.

\bibliography{Ref}

\section*{Acknowledgments}

F. A. acknowledges the Inter University Centre for Astronomy and Astrophysics (IUCAA), Pune, India for granting visiting associateship. \.{I}.~S. thanks T\"{U}B\.{I}TAK, ANKOS, and SCOAP3 for funding and acknowledges the networking support from COST Actions CA22113, CA21106, and CA23130.

\section*{Funding Statement}

No funds have been received for this manuscript.

\section*{Data Availability Statement}

In this study, no new data was generated or analyzed.

\end{document}